\documentclass[9.5pt,journal,compsoc]{IEEEtran}        % review (journal style)
\usepackage{CJKutf8}
\usepackage{hyperref}
\usepackage{graphicx}
\graphicspath{{figures/}{pictures/}{images/}{./}} % where to search for the images

\usepackage[section]{placeins}
\usepackage{graphicx}
\usepackage{amsmath}
\usepackage{microtype}                 % use micro-typography (slightly more compact, better to read)
\usepackage{textcomp}                  % use better special symbols
\usepackage{cite}                      % needed to automatically sort the references
\usepackage{tabu}                      % only used for the table example
\usepackage{booktabs}                  % only used for the table example
\usepackage{bm}
\usepackage{xcolor}
\usepackage{subfigure}
\usepackage{relsize}
\usepackage{threeparttable}
\usepackage{soul}
\usepackage{xspace}

\usepackage{wrapfig}
\usepackage{enumitem}
\def\etal{{\textit{et~al.}}}

\newcommand\od[1]{{{#1}}}

\newcommand{\yh}[1]{{#1}}

\begin{document}
% Titles are generally capitali,zed except for words such as a, an, and, as,
%,,,,,,,,,,\title{t-FDP: Improving Force-Directed Graph Layouts with Heavy-tailed Forces}
\title{Force-directed graph layouts revisited: a new force based on the t-Distribution}
% Revisiting,,,

%\title{Categorical Data Colorization for Juxtaposed Views based Comparison}
%Color Design for Comparing Multiple Juxtaposed Datasets}
%\author{}Colorization for Categorical Data Colorizing
%% Abstract section.
\author{Fahai Zhong, Mingliang Xue, Jian Zhang, Fan Zhang, Rui Ban, Oliver Deussen and Yunhai Wang
	% <-this % stops a space
	\IEEEcompsocitemizethanks{
		\IEEEcompsocthanksitem Fahai Zhong, Mingliang Xue, Yunhai Wang are with the Department of Computer Science, Shandong University, China. E-mail: \{zhongfahai, xml95007, cloudseawang\}@gmail.com
		\IEEEcompsocthanksitem Jian Zhang is with Computer Network Information Center, Chinese Academy of Sciences, Beijing, China. E-mail: zhangjian@sccas.cn
		\IEEEcompsocthanksitem Fan Zhang is with School of Computer Science and Technology, SDTBU, China. E-mail: zhangfan@sdtbu.edu.cn
		\IEEEcompsocthanksitem Rui Ban is with Intelligent Network Design Institute, CITC, Beijing, China. E-mail: banrui1@chinaunicom.cn
		\IEEEcompsocthanksitem Oliver Deussen is with  Computer and Information Science, University of Konstanz, Konstanz, Germany. E-mail: oliver.deussen@uni-konstanz.de
		\IEEEcompsocthanksitem Yunhai Wang is the corresponding author
	}% <-this % stops an unwanted space
}

\IEEEtitleabstractindextext{%
	\vspace{-2mm}
	\begin{abstract}
		%Graphs are an ubiquitous form of data, whenever objects and their relations have to be represented. Shown as node-link diagrams, they are intuitive and support many analysis tasks. Being high-dimensional data, their dimensionality has to be reduced for display, a task for which force-directed placement (FDP) methods are often applied. The graph is modeled as a physical system of bodies with forces acting between them, the layout is computed by minimizing the overall energy of the system when compressed to two dimensions.  This physical metaphor has inherent drawbacks such as distorting local neighbourhoods. We therefore argue that it is necessary to deviate from it and propose the t-FDP model, a force-directed placement method based on a novel, bounded short-range force defined by t-distribution. Our formulation is flexible, exerts limited repulsive forces on nearby nodes and can be adapted separately in its short- and long-range effects. This yields to a better neighborhood preservation while maintaining the overall structure of the results.
%Our efficient implementation using a Fast Fourier Transform is one magnitude faster than state-of-the-art methods and two orders faster on the GPU, enabling us to perform interactive parameter optimization.
In this paper, we propose the t-FDP model, a force-directed placement method based on a novel bounded short-range force (t-force) \yh{defined by Student's t-distribution}.
Our formulation is flexible, exerts limited repulsive forces for nearby nodes and can be adapted separately in its short- and long-range effects.
Using such forces in force-directed graph layouts yields better neighborhood preservation than current methods, while maintaining low stress errors.
%most problems with
%The model allows us to create node-link diagrams of high quality, an efficient implementation using a Fast Fourier Transform on the GPU enables us to perform interactive parameter optimization. We demonstrate the quality of our approach by numerical evaluation against state-of-the-art approaches and extensions for interactive exploration.
Our efficient implementation using a Fast Fourier Transform is \yh{one order of magnitude} faster than state-of-the-art methods and two orders faster on the GPU, enabling us to perform parameter tuning by globally and locally adjusting the t-force \yh{in real-time for complex graphs.}
We demonstrate the quality of our approach by numerical evaluation against state-of-the-art approaches and extensions for interactive exploration.

 \vspace{-2mm}
	\end{abstract}

	}

\maketitle

\IEEEdisplaynontitleabstractindextext


% \fi
\IEEEpeerreviewmaketitle

\section{Introduction}
\begin{figure*}[t]
    \centering
    \vspace{-4mm}
    \includegraphics[width=0.97\linewidth]{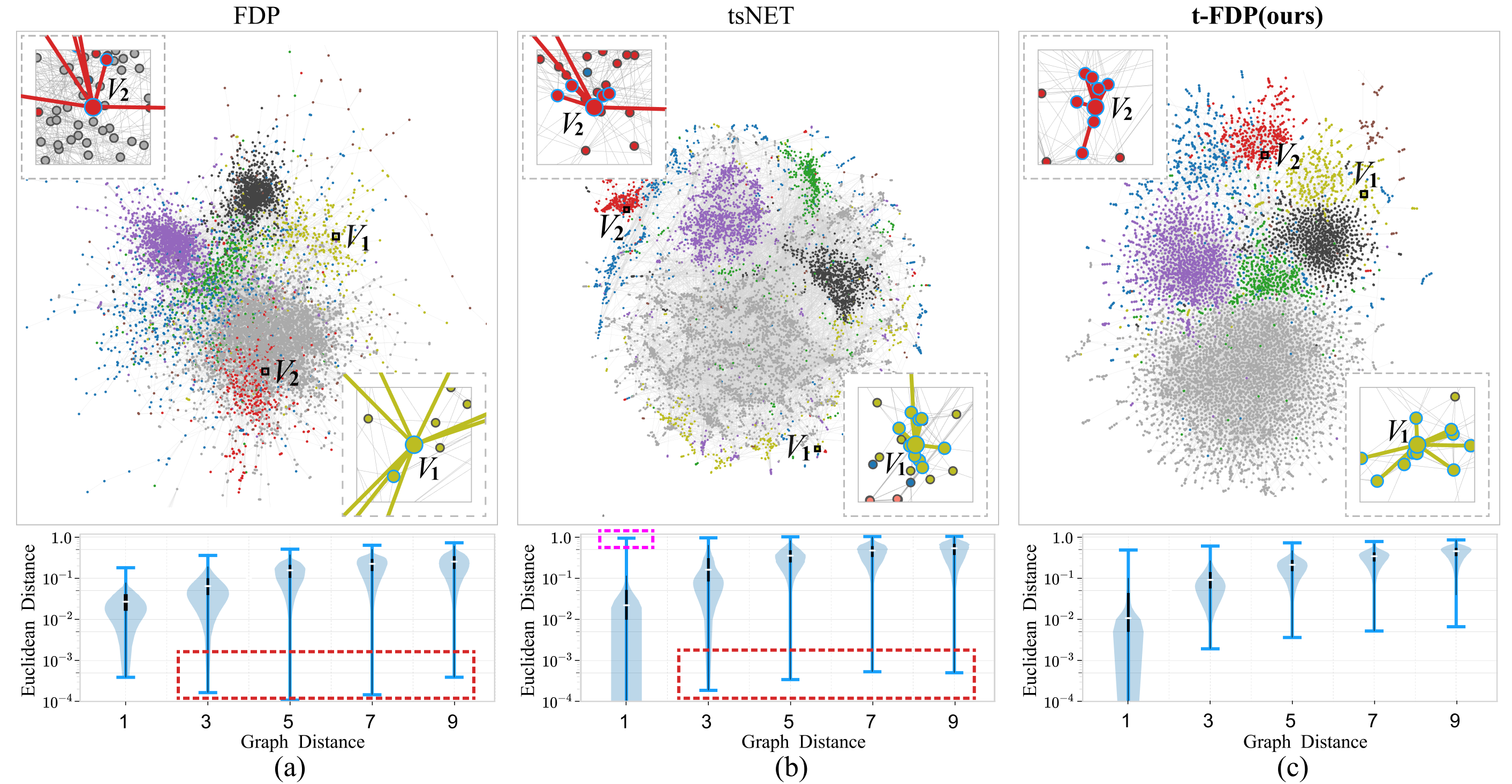}\vspace{-4mm}
    \caption{
        Comparison of our technique (t-FDP) with standard  force-directed placement (FDP) \cite{fruchterman1991graph} and tsNET~\cite{kruiger2017graph}, a state-of-the-art neighborhood embedding-based approach on a labeled co-author network by summarizing the normalized Euclidean distances of graph nodes in
        layout space with violin plots shown in the bottom of (a,b,c).
        While FDP places nodes close by that have large graph distances (the red box on the  violin plots in a),
        tsNET avoids this to some extent (the red  box on the violin plots in b) but often keeps nodes with small graph distances far from each other (the pink box on the violin plots in b). t-FDP in (c) is able to reach both, short layout distances for nearby points in data space and large ones for distant data points. Therefore it separates all eight classes clearly, while the other methods mix them in part.
        Two example neighborhoods show this: in (a) the neighbours of $V_1$ do not have connections to $V_1$ and the graph neighbours of $V_2$ are mostly far away.  tsNET is better in keeping the local neighborhood of $V_1$, but the neighborhood of $V_2$ has large graph distances. Both cases have a better neighborhood preservation with t-FDP.
    }
    \vspace{-5mm}
    \label{fig:teaser}
\end{figure*}

\IEEEPARstart{G}{raphs} are a ubiquitous form of data, whenever objects and their relations have to be represented; lots of effort has been spent on finding good ways to visualize them. Arguably one of the most prevalent representations for graphs are node-link diagrams, which are intuitive and efficient in supporting various graph analysis tasks~\cite{lee2006task}. %by revealing local and global structures.

Despite the existence of many other techniques, force-directed placement (FDP) methods~\cite{tamassia2013handbook} have gained a lot of attention for automatically layouting node-link diagrams.  %due to their naturalness and intuitive manipulation possibilities.
Here, a graph is modeled as a physical system of bodies with forces acting between them, the layout is computed by minimizing the overall energy of the system. %we cannot say it is an optimal solution, only the result of an optimization
While some variants of forces have been proposed \cite{tamassia2013handbook}, the spring-electric model~\cite{eades84Heuristic,fruchterman1991graph} is the most popular method. It assigns spring-like attractive forces between connected nodes and repulsive electrical forces between all node pairs to keep them at a distance.
For most graphs, the method creates layouts with compelling performance~\cite{gansner2012maxent}, especially when applying its multi-level version, the scalable force-directed placement (SFDP)~\cite{hu2005efficient}.
%aims to place neighboring nodes close to each other while ensuring that nodes are not too close to each other
Since it is intuitive and easy to implement, the method is a key component in many visualization systems (e.g. Gephi~\cite{bastian2009gephi} and D3 \cite{Bostock11DDD}). %are based on such layouts.
%While many methods have been proposed to improve the scalability of the spring-electric model~\cite{hu2005efficient,yunis2012scalable,gove2019random}, to our knowledge no study was presented to validate that the spring-electric model is most appropriate for modeling graph layout problems.

%especially the multi-level version, scalable force directed placement (SFDP) method ~\cite{hu2005efficient}.
%Much effort has been devoted to improving the scalability of the spring-electric model in past decades~\cite{hu2005efficient,yunis2012scalable,gove2019random}.
%To further improve the local structure preservation, neighbored embedding based  dimensionality reduction (DR) techniques have been leveraged for graph layouts.
%In this paper, we revisit the spring-electric model and find that its underlying physical laws cannot properly capture local neighborhoods and cluster structures in graphs.
%A major problem with spring-electric models is that they cannot always properly capture local neighborhoods and cluster structures (see Fig.~\ref{fig:teaser}(a)).
%
A typical problem of spring-electric models is that large repulsive electric forces are exerted on each nearby pair of connected nodes, but only small attractive spring forces.
In a physical simulation this is reasonable since repelling contact forces for real bodies should be extremely large. However, such forces might destroy graph structures by \od{preventing connected nodes in the graph from being neighbors in the layout} 
% preventing connected nodes in the layout from being neighbors in the graph 
(see the bottom statistics in Fig.~\ref{fig:teaser}(a)).
Although  recently proposed graph layout methods such as tsNET~\cite{kruiger2017graph}  allow to better preserve \od{2}-ring neighborhoods, they fail in efficiently maintaining low stress errors and 1-ring neighborhoods (see Fig.~\ref{fig:teaser}(b)).
DRGraph~\cite{zhu2020drgraph} improves the scalability of tsNET but has a similar issue with yielding large stress errors. %overall structures.

%, its improved version, and is roughly two times faster than SFDP.
%% but fails in efficiently maintaining the overall graph structure.  %Moreover, their results are hard to interpret due to the inherent complexity of the optimizations used for t-SNE~\cite{chatzimparmpas2020t}.

In this paper, we revisit the spring-electric model and propose a new heavily-tailed force
based on the t-distribution (referred to as t-force), %based on the Student's t-distribution energy
for better maintaining local  graph structures while keeping the intuitiveness and simplicity of the model.
Deviating from traditional spring-electric models gives us the flexibility to freely explore the interplay between attractive and repulsive forces.
This is achieved by dividing forces into their short- and long-range components and improving their short-range aspects.
The t-force has an upper bound at short-range and behaves similar to the electrical repulsive force at long-ranges. The attractive force at short-range is kept unaltered.
%can be adjusted in its maximum value and heaviness.
%As mentioned above, we found that the gradient of the Student's t-distribution~\cite{pearson1895x} (simply t-distribution) works very well as this force, we will refer to it as \textit{t-force}.
%This force reaches its maximum magnitude at a short distance between two nodes and then smoothly decays to zero as the distance decreases or increases.
%For distances close to infinity, this force is equivalent to an electric force; when the distance reaches zero, the force magnitude behaves like a spring.
%Both, maximal force magnitude and tail heaviness can be adjusted by a single parameter.
By using t-forces to define the repulsive force within FDP and combining t-forces with spring forces as the attractive forces, we are able to compose a new t-force-based FDP model (\textit{t-FDP} for short), which allows us at the same time better preserving graph neighborhoods, distances of nodes and clusters  (see Fig.~\ref{fig:teaser}(c)).

Being one of two aspects of the t-SNE gradient, the t-force is  similar to the repulsive force of the t-SNE model~\cite{vandermaaten2008visualizing}, a widely used method for dimensionality reduction. This allows us  to employ an interpolation-based acceleration strategy for t-SNE using the fast Fourier transformation~\cite{linderman2019fast} which can be implemented on the GPU.
%\od{Our t-force has a similar form to the repulsive force of the t-SNE model~\cite{vandermaaten2008visualizing}, since it is one of two parts of the t-SNE gradient.} Therefore, we are able to employ the recently proposed  interpolation-based acceleration strategy for t-SNE using the fast Fourier transformation (FFT) ~\cite{linderman2019fast} to implement our t-FDP model efficiently.
Due to the parallel nature of the FFT, our whole model can be efficiently implemented on GPUs and this allows us to perform interactive parameter optimization.
To enable a wide range of users to use our method, we provide an implementation as a drop-in force for the  ``d3-force'' library.
% In order to promote our method to a wide range of users,

We evaluated our approach using 50 graphs with node numbers ranging from  72 to 4 million and edge numbers ranging from 75 to 100 million. We quantitatively measured the quality of our results using various structural and readability metrics. For the tested data, our method performs similarly or even outperforms all existing methods in all metrics except the stress error, where our method is the second-best behind the stress model.
Regarding performance, the CPU version of t-FDP is one order of magnitude faster than SFDP and DRGraph on a desktop computer with Intel i7-8700 CPU, while reducing the memory size considerably. Our GPU-based CUDA implementation is able to further improve performance by another order of magnitude.

In addition, we present two extensions of t-FDP for better exploring graph structures.
First, we show how we reveal structures at different scales by re-applying our force to existing layouts with repulsive forces of different magnitudes. Second, we allow users to explore nodes of interest with fisheye-like visualizations by interactively changing  force parameters. Since our method is very fast, interactivity can be maintained even for larger graphs.
%respectively. One is to increase the weight of the repulsive t-force for enlarging the repulsion between nearby nodes, the other is to decrease the maximum magnitude of the repulsive t-force for moving connected nodes closer together in the visual representation. Initializing these two extensions with the results of the t-FDP model can quickly generate a variety of desired results.We demonstrate the effectiveness with three case studies.

In summary, our main contributions are as follows:
\begin{itemize}
    \item we revisit the spring-electric model and show that it does not properly capture local neighborhoods and cluster structures in graphs;

    \item we propose a new short-range force based on the t-distribution and use it to compose a new FDP model that overcomes drawbacks of existing FDP models in neighborhood preservation; and

    \item we provide an FFT-based fast implementation\footnote{ \url{https://github.com/Ideas-Laboratory/t-fdp}} and \yh{an interactive demo\footnote{\url{https://t-fdp.github.io}} of} our method,
          and demonstrate its effectiveness through a quantitative evaluation and extensions.
\end{itemize}

\section{Related Work}
Related work falls into two categories: force-directed graph layouts and the  t-distribution and its applications in graph layout.

% points on edge. For repulsion, FDP means that all points are repulsed by electrons, while MaxEnt means that two points on the edge are repulsed by springs and other points are repulsed by electrons. And Stress Majorization is the spring attraction and repulsive force related to the ideal length dij between all points.

%Various graph layout techniques have been proposed for computing node coordinates (typically 2D) for a given undirected graph. A complete survey of these techniques is beyond the scope of this paper; we refer the reader to~\cite{tamassia2013handbook,von2011visual}. Here, we restrict our discussion to the techniques that are particularly relevant to our work: force-directed and dimensionality reduction layout methods and the approximation methods for acceleration.

\subsection{Force-directed Graph Layouts}
Since their first proposition by  Tutte~\cite{tutte1963draw}, many variants of force-directed layouts have been developed~\cite{tamassia2013handbook,chen2019survey}. Among them,
spring-electric and  stress models are two most popular categories.

%Tutte's barycentric method~\cite{tutte1963draw}, which can obtain a cross-free layout for 3-connected planar graph. To handel general graphs,
\vspace{1mm}
\noindent\textbf{Spring-electric model}.
By representing nodes as ring-like joints and edges as springs,  the spring-electric model~\cite{eades84Heuristic,fruchterman1991graph} uses attractive forces to pull adjacent nodes together and repulsive forces between all nodes to repel them. When the energy of the system reaches a minimum, it finds the optimal layout. To avoid strong long-range forces, Eades~\cite{eades84Heuristic} introduced
logarithmic spring forces and repulsive  forces inverse to the squared distance.
To produce layouts with uniform edge lengths, Fruchterman and Reingold~\cite{fruchterman1991graph} (FR) redefine attractive forces  proportional to the squared distance and repulsive forces reciprocal to the distance.
Hu et al.~\cite{hu2005efficient} model repulsive forces as power functions of the form $f(d) = d^{-p}$ with $p$ being any positive integer
and find that forces inversely related to the squared distance work well for most graphs. Noack~\cite{noack2007energy} generalizes this idea to a LinLog energy model which uses constant attractive forces and repulsive forces reciprocal to the distance, yielding well-separated clusters for their tested graphs.

As a compromise between LinLog and FR model, ForceAtlas2~\cite{jacomy2014forceatlas2}
sets  attractive  forces proportional and repulsive forces
reciprocal to the distance for better showing local neighborhoods and cluster structures.
However, as pointed out by Fruchterman and Reingold~\cite{fruchterman1991graph}, such a configuration might work poorly for complex graphs, because strong short-range repulsive forces often let the corresponding optimization be trapped in local minima. To alleviate this issue, Kumar \etal~\cite{NodeInterleaving} introduce a heuristic method that imposes an upper bound on the repulsive forces as a stopping condition for visualizing directed acyclic graphs.
In contrast, short-range t-forces in our t-FDP model allow us to largely alleviate this issue for general graphs (see Section~\ref{sec:tfdpm}).

To visualize large graphs, various multilevel approaches have been employed for improving scalability. For example, the Barnes-Hut technique~\cite{barnes86Hierarchical}  and the fast multipole method~\cite{greengard1987rapid} are often used for approximating  long-range repulsive forces~\cite{hu2005efficient}, having an overall time complexity of $O(n\log n)$ for $n$ nodes. Since the Barnes-Hut approximation is easy to implement, it is adopted by many FDP algorithms such as SFDP~\cite{hu2005efficient} and ForceAtlas2~\cite{jacomy2014forceatlas2}.
Random vertex sampling was proposed by Gove~\cite{gove2019random} for computing repulsive forces. It  generates similar layouts as Barnes-Hut and FMM, but with a time complexity of  $O(n)$. Based on FFT acceleration, our t-FDP model also shows a runtime of $O(n)$,  but the resulting layouts  maintain \yh{lower stress errors.}

\vspace{1mm}
\noindent\textbf{Stress model}.
Alternatively, stress models, originally proposed by Kamada and Kawai~\cite{kamada1989algorithm} associate a spring between each pair of nodes with an ideal length proportional to the length of the shortest path between them. In doing so, this model can yield a layout with a much better global structure than a spring-electric model. %To shorten running time, advanced efficient optimization methods~\cite{gansner2004graph} and sparse/low-rank approximation stress models were proposed~\cite{ortmann2016sparse,khoury2012drawing}.
However, stress models pay more attention to distant node pairs, resulting in typically poor preservation of local structures.

To preserve such local structures better, local versions of stress models~\cite{koren2008binary,chen2009local,gansner2012maxent,ConstrainedSM} have been proposed. The maxent-stress model~\cite{gansner2012maxent} applies stress only on edges within a specified length, but electric repulsion forces to all nodes. This reduces the computational costs for the shortest paths of all node pairs.
In contrast to this, t-FDP  aims at  improving spring-electric models by better balancing  the representation of local and global structures.
Our experimental results show that it performs better than the maxent-stress model for representing global structures while being significantly faster.

\subsection{t-Distribution and its Applications for Graph Layout}
The t-distribution~\cite{pearson1895x} is a symmetric bell-shaped distribution with heavier tails compared with the Gaussian distribution.
%\begin{wrapfigure}{r}{0.3\linewidth}
%    % \vspace{-10pt}
%    \includegraphics[width=0.98\linewidth]{distribution.png}\vspace{-4mm}
%    \caption{$f(x)$ for different $v$.} \vspace{-4mm}
%    \label{fig:distributioncurve}
%\end{wrapfigure}
%\vspace{8mm}
The function is defined by:
%% distribution2 is a wider version.
% \begin{figure}[t]
%     \centering
%     \includegraphics[width=0.8\linewidth]{distribution2.pdf}\vspace{-3mm}
%     \caption{The t-distribution for different $v$.}
%     \label{fig:distribution}\vspace{-8mm}
% \end{figure}

%\begin{align}
$$
    f(x) = \left(1+\frac{x^2}{2v-1}\right)^{-v}, \nonumber
$$
%\end{align}
where $v$ is the degree of freedom. With increasing $v$ the distribution becomes closer to the Gaussian distribution. %(see Fig. \ref{fig:distributioncurve}).
% \odc{we might show a little image comparing Gaussian and t-distribution}

By employing such a distribution to model the similarity between two points in the embedding space, the neighborhood embedding t-SNE~\cite{vandermaaten2008visualizing} not only greatly alleviates the crowding problem of dimensionality reduction (DR) but also efficiently preserves local structures.

A number of  t-distribution based DR methods have been proposed, such as LargeVis~\cite{tang2016visualizing}, UMAP~\cite{mcinnes2018umap} and TriMap~\cite{amid2019trimap} for preserving more global structures.
On the other hand, acceleration strategies were developed for improving the scalability of t-SNE, such as the Barnes-Hut approximation~\cite{van2014accelerating}, GPU-based texture splatting~\cite{pezzotti2019gpgpu} and interpolation-based FFT~\cite{linderman2019fast}.  Among them, the interpolation-based FFT method can achieve a near-linear complexity by using the fast Fourier transform to approximate the repulsive force, which is one term of the negative gradient.

Recently, t-distribution based DR methods have been leveraged for graph layout.
A representative approach is tsNET~\cite{kruiger2017graph}, which utilizes a customized t-SNE for capturing local structures. Compared to the stress model based on multidimensional scaling MDS~\cite{kruskal1978multidimensional}, tsNET can generate high-quality layouts for a wider variety of graphs. To handle large graphs, Zhu et al. propose DRGraph~\cite{zhu2020drgraph} that %leverages three strategies used by UMAP~\cite{mcinnes2018umap} to
further improves the scalability of tsNET by using  negative sampling for gradient estimation.

In a similar spirit, we use the t-distribution  to define the bounded short-range force in our t-FDP model. Furthermore, this formation enables us to employ the FFT~\cite{linderman2019fast} for improving the scalability, which is about one order of magnitude faster than DRGraph on the CPU and two orders faster on the GPU.

\section{Method}
Given an undirected graph with $n$ nodes and $m$ edges, the main goal of graph layout is to compute a low-dimensional position for each node that meets a set of structural and aesthetic criteria including evenly distributed nodes, uniform edge lengths, and  reflecting symmetry. However, force-directed layout algorithms and especially spring-electric models do not explicitly strive for these goals but are solely based on two design principles proposed by Fruchterman and Reingold~\cite{fruchterman1991graph}:
\begin{itemize}[label=(\roman*),nosep]
	\item[\textbf{P1:}] Nodes connected by an edge should be drawn close to each other; and
	\item[\textbf{P2:}] Nodes should not be drawn too close to each other in general.
\end{itemize}
These two principles, however, cannot ensure that connected nodes become nearest neighbors in the layout for graphs  with more than two nodes (see Fig.~\ref{fig:teaser}(a)). This results in losing important local graph structures.
To address this issue, we propose a third design principle:
\begin{itemize}[label=(\roman*),nosep]
	\item[\textbf{P3:}] Nodes connected by an edge should be drawn closer to each other than unconnected nodes.
\end{itemize}
This will enhance local structures and represent local connection patterns more faithfully.
%In the following, we briefly review how the spring-electric model satisfies these principles, but then we have to discuss its limitations.

\subsection{Revisiting Spring-electric Models}
%By representing vertices and edges as atomic particles and springs,
To meet \textbf{P1} and \textbf{P2}, the spring-electric model employs attractive spring forces $f^a$ to pull connected nodes together and electric repulsive forces $f^r$ to repel nodes from each other.
So far, different functions have been used for defining attraction and repulsion forces~\cite{hu2005efficient,jacomy2014forceatlas2} and many of them can be written in the form of power functions:
\begin{align}
	F^{a}(i,j) & = \alpha||\mathbf{x}_i-\mathbf{x}_j||^p,           \quad i\leftrightarrow j ,
	\label{eq:powerattractive}                                                                   \\
	F^{r}(i,j) & = -||\mathbf{x}_i-\mathbf{x}_j||^{-q}, \quad i\neq j, \label{eq:powerreplusive}
\end{align}
where $\mathbf{x}_i$ is the 2D position of the graph node $i$ in the layout space, $\alpha$  is the weight, and $p$ and $q$ have to be non-negative.
The larger $p$, the stronger the long-range attractive force, while the larger  $q$, the weaker the long-range repulsive force.
%The default values of $(p,q)$ are (1,1) in  many visualization libraries~(e.g.,  \cite{Bostock11DDD,bastian2009gephi}).
At each time step, the resultant force will move the node until convergence is reached.
%and $k$ is the optimal distance between vertices~\cite{fruchterman1991graph}.
%Hu et al.~\cite{hu2005efficient} show that %adjusting $k$ and $w$ merely scales the layout, while

The layout quality is mainly influenced by $p$ and $q$.
By using the exponents $(2,1)$ for $p$ and $q$, respectively, the FR model~\cite{fruchterman1991graph} yields layouts with uniform edge lengths. The LinLog model~\cite{noack2007energy}  uses the exponents $(0,1)$ for better revealing clusters in the graph.
Recently, Jacomy et al.~\cite{jacomy2014forceatlas2}
suggest to use the exponents $(1,1)$ as an intermediate model between these two models, these values are also set as default in many visualization libraries or systems such as GraphViz~\cite{gansner2009drawing} or D3~\cite{Bostock11DDD}.

%Many methods exist to overcome local minima and improving scalability. However,

%
%
%\begin{figure}[ht]
%    \centering
%    \includegraphics[width=0.98\linewidth]{fig2}
%    % \includegraphics[width=0.48\linewidth]{tdis.pdf}
%    % \subfigure[]{\includegraphics[width=0.48\linewidth]{repforce_current.pdf}}
%    \caption{The current FDP approach causes untrustworthy neighbors and weakens clusters. (a) the same initial layout. (b) The forces of the current(left) and our FDP method (right). (c) The layout after five iterations with the current FDP method. (d)  The layout after five iterations with our FDP method.}
%    \label{fig:current_FDP}
%\end{figure}

\begin{figure}[t]
	\centering
	% \subfigure[]{\includegraphics[width=0.48\linewidth]{tdis.pdf}}
	\includegraphics[width=0.92\linewidth]{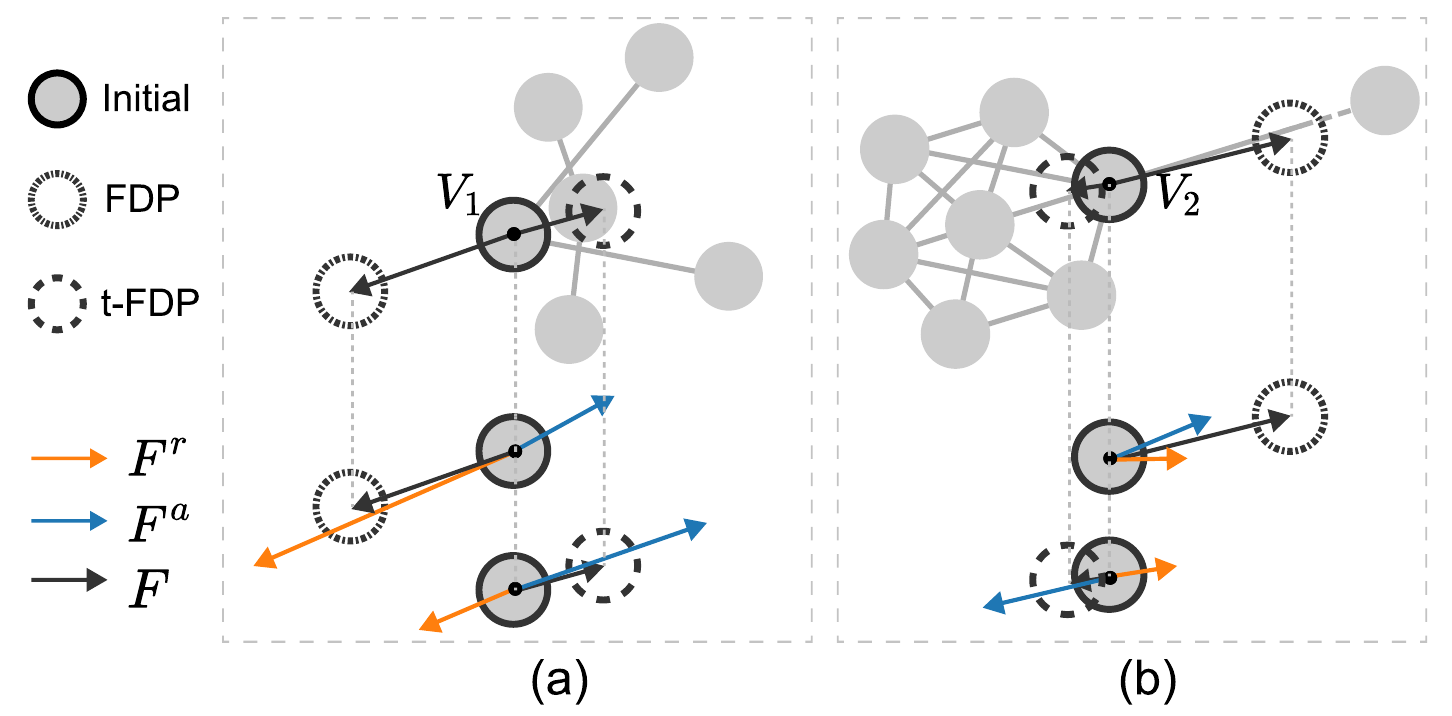} \vspace{-4mm}
	% \subfigure[]{\includegraphics[width=0.48\linewidth]{repforce_current.pdf}}
	\caption{Two problematic cases of traditional FDP: (a) three nearby non-neighbouring nodes exert large repulsive forces to push $V_1$ out of its cluster; (b) a strong long-range attractive force pulls $V_2$ out of a local cluster.  t-FDP avoids both cases.} \vspace{-3mm}
	\label{fig:tdisforce}
\end{figure}

\vspace{2mm}
\noindent\textbf{Drawbacks}.
%\label{sec:limits}
Attractive spring and repulsive electric forces allow to meet \textbf{P1} and \textbf{P2} for small and simple graphs. However, for large graphs, such forces based on power functions might not be able to adequately represent graph neighborhoods and cluster structures.
%Few studies investigate if such forces are appropriate for solving graph layout problems. The only exception
%was performed by Eades~\cite{eades84Heuristic} who found that linear spring forces (p = 1) are too strong on the long range and  he introduced logarithmic spring forces. Yet, such forces are quite costly to compute.
In the following, we describe the drawbacks of classical FDP detailing the  two problematic cases shown in Fig.~\ref{fig:tdisforce}.

First, as mentioned above, repulsive forces tend to become extremely large when the Euclidean distance between two nodes approaches zero. As a result, closely connected nodes might be pushed away from each other (see $V_1$ in Fig.~\ref{fig:tdisforce}(a)). In doing so, the resulting Euclidean distances often do not correlate with graph-theoretical distances (see Fig.~\ref{fig:teaser}(a)).
%most  neighboring points of a node are only loosely connected to it after a few iterations and the formation of a clusters is suppressed. The boxplots shown in the bottom of As Fig.~\ref{fig:teaser}(a) summarize the normalized Euclidean distances between connected nodes in the corresponding FDP layout, where the closest nodes are not the connected nodes in the graph.
This is counter-intuitive for graph exploration.

Second, long-range attraction forces that are too strong might break a cluster into multiple smaller pieces. As shown in Fig.~\ref{fig:tdisforce}(b),  node $V_2$ is affected by a strong long-range attraction force from an outlier and moved out of the main cluster. Reducing long-range attraction forces or increasing short-range attraction might reduce this issue. However, adjusting the corresponding coefficients in the spring-electric model will always change both, long- and short-range forces at the same time.

%In addition, combining both forces might cause the optimization easily to be trapped in a local minimum. As shown in Fig.~\ref{fig:tdisforce}(a), the repulsive force exerted by three non-neighboring vertices on vertex $V_1$ is larger than the attractive forces exerted by its neighboring vertices.
%As a result, $V_1$ cannot move towards its connected vertex. After a few iterations, the cluster, which is formed by the connected nodes might be split into multiple sub-clusters.

Although the above two issues can be alleviated by assigning different coefficients to the forces at different levels as possible in multilevel methods~\cite{walshaw2000multilevel,hachul2004drawing,hu2005efficient}, power-function-based forces inherently suffer from exerting extremely large repulsive forces and small attractive forces for short-ranges.
On the other hand, small repulsive and large attractive forces for long-ranges help to reduce long edges, resulting in a layout with reasonably global structures.
\od{Thus, we propose our new model that decouples these effects and allows for improving repulsive and attractive forces at short-ranges, while keeping the characteristics of the traditional FDP models at long-range.}

% ensure a layout for better revealing graph graph structures.
% while providing a FFT based algorithm for fast and simple implementation (Section 5).

%Fig.~\ref{fig:small_case} shows a small example. This example demonstrates how the current FDP approach can lead not only to weakened clusters (yellow vertices) and untrustworthy neighbours (green vertices in the circle), but also to incorrect locations between clusters (black vertices, orange and blue clusters).

%\begin{figure}[t]
%    \centering
%    \includegraphics[width=0.98\linewidth]{tforce_smallcase_2}
%    % \includegraphics[width=0.48\linewidth]{tdis.pdf}
%    % \subfigure[]{\includegraphics[width=0.48\linewidth]{repforce_current.pdf}}
%    \caption{\od{(a) Small graph layout by the Fruchtermann-Reingold algorithm. (b) When replacing the repulsive force with the t-force, local structures are better represented. (c) Further combining the t-force with a spring force results in more compact clusters. (?)}}
%    \label{fig:small_case}
%\end{figure}

\subsection{The t-Force}

%As the graph size increases, it is almost impossible to show all nodes without overlap on a limited display surface. Hence, more attention needs to be paid for trustworthily representing local neighborhoods and cluster structures. To do so, we introduce an additional design principle:
%\begin{enumerate}[label=(\roman*),nosep]
%    \item \textbf{P3:} Vertices connected by an edge should be drawn closer than unconnected vertices.
%\end{enumerate}

One straightforward way to avoid extremely large repulsive forces would be to define a constant force value  when two nodes approach each other.
However, only manipulating repulsive forces with a constant cannot meet \textbf{P3}, because it is hard to find a good balance to the spring-like attractive forces.
%However, power function based repulsive forces are too large  at the short range that might disturb other structures, while the attractive forces are too large at the long range.
Therefore, it is better to increase  attraction and reduce repulsion at short-ranges so that neighbors are able to move closer together.
%In addition, \textbf{P2} requires that the repulsive forces  at very short ranges should be larger than the attractive forces to avoid collapsing local structures.
To do so, \od{we propose the t-force as a new short-range force} that serves as two roles: a new repulsive force and a component of the attractive force for short-ranges.

As a repulsive force, it should be short-range in nature, and thus we require it to be weak at long ranges similar to power-function-based repulsive forces.
Moreover, it should have an upper bound at short ranges to avoid extremely large contact forces and improve optimization stability.
Finally, to be used as an additional component of attractive forces at short ranges, we further require it to have similar behavior as a linear spring force, which is quite strong compared to other forces as shown by Eades~\cite{eades84Heuristic}.
Suppose the Euclidean distance between two graph nodes is $d= \|\mathbf{x}_{i}-\mathbf{x}_{j}\|$, the force $f(d)$ should satisfy the following three requirements:
\begin{itemize}
    \setlength\itemsep{1mm}
    \item \textbf{R1:}  $\exists \varsigma>0\mbox{  } \mathrm{s.t.} \mbox{  }$ $0 < f(d) \leq \varsigma, \; \forall d>0$;
          \vspace{-1mm}
    \item \textbf{R2:} $f(d) \sim d^{-q} \quad\mathrm{as~}  d \to \infty , \mbox{ } \mathrm{where}\mbox{ }\, q>0$; and %$ \forall q>0, \lim\limits_{d \to \infty} f(d) = d^{-q}$;  and
          \vspace{-1mm}
    \item \textbf{R3:} $ f(d) \sim d  \quad\mathrm{as~}  {d \to 0}$.
\end{itemize}
In other words, such a force has an upper bound $\varsigma$ for short distances between two nodes and smoothly reduces to zero as the distance decreases with different possible decay rates.
%(\textbf{R1}) and then \textbf{R2} indicates that this force is equivalent to a long-range electric force when distances close to infinity, while \textbf{R3} indicates that the force magnitude behaves like a short-range spring when the distance reaches zero.

%defined by Eq.~\ref{eq:powerattractive}  and Eq.~\ref{eq:powerreplusive}, respectively.

%However, the space for searching the commentation of multiple proper functions is huge.

%To address this challenge, we propose a new definition of attraction and repulsion forces using the t-distribution that helps to form a trustworthy force directed placement (t-FDP) model. Fig.~\ref{} shows an example, where the layout generated by our t-FDP model greatly capture the local neighborhood and cluster structures.

%\subsection{t-Distribution based Forces}
After investigating a number of functions, we found the following t-distribution based force to meet our requirements. This does not mean that no other possible functions exist, but this one seems to be very simple and provides all required aspects:
\begin{equation}
    f(d) = \frac{2\tau \tilde{\varphi} d}{(1+ \tau d^2)^{\tilde{\varphi}+1}} , \label{eq:generalizetforce}
\end{equation}
where $\tau$ and $\tilde \varphi$ are constants not less than zero. %The formula is the negative derivative of a generalized t-distribution~\cite{yang2009heavy}:
%\begin{equation}
%    g(x) = \frac{1}{(1+\tau x^2)^\varphi}.
%    \label{eq:t-distribution}
%\end{equation}
%For $\varphi$ and $\tau$ equal to 1, this is the standard t-distribution used by t-SNE.
For simplicity, we set $\tau$ to 1 and simplify the notation of Eq.~\ref{eq:generalizetforce} to the following function: %of $f(x)$ to [0,1]. Thereby, we have
\begin{align}
    f(d) & = \frac{d}{(1+ d^2)^\varphi}
    \label{eq:forcefunction}
\end{align}
with exponent $\varphi\geq 1$.

The function $f(d)$ approaches 1 when $|d|$ is close to zero and gradually decreases to zero as $|d|$ increases. We refer to the heavy-tailed force defined by  Eq.~\ref{eq:forcefunction} as \textit{t-force}.
Fig.~\ref{fig:functiongraphs}(a) shows $f(d)$ for three different values of $\varphi$. The maximum  is always smaller than 1 and located in the $d$-range of [0,1],  $f(d)$ is heavily influenced by $\varphi$. The smaller  $\varphi$, the larger the maximum and the heavier its tail. % whose energy is the t-distribution of ~\eqref{eq:t-distribution}.
No matter what $\varphi$ is, the function has an upper bound of $\varsigma$.

%By directly taking $f(x)$ as the repulsive force, \textbf{R1} can be met.
\subsection{The t-FDP model}\label{sec:tfdpm}
Using the t-force, we now define the repulsive and attractive forces of our force-based graph layout as: %
\begin{footnotesize}
    \begin{align}
        F^r(i,j)
         & = -\frac{||\mathbf{x}_{i}-\mathbf{x}_{j}||}{(1+||\mathbf{x}_{i}-\mathbf{x}_{j}||^2)^ {\gamma}} \frac{\mathbf{x}_{i}-\mathbf{x}_{j}}{||\mathbf{x}_{i}-\mathbf{x}_{j}||}, \quad i\neq j,
        \label{eq:replusive}
        \\
        F^a(i,j)
         & = \left(||\mathbf{x}_{i} - \mathbf{x}_{j}|| +
        \frac{\beta||\mathbf{x}_{i}-\mathbf{x}_{j}||}{1+||\mathbf{x}_{i}-\mathbf{x}_{j}||^2}\right)
        \frac{\mathbf{x}_{i}-\mathbf{x}_{j}}{||\mathbf{x}_{i}-\mathbf{x}_{j}||} ,\quad i\leftrightarrow j ,
        \label{eq:newforce}
    \end{align}
\end{footnotesize}
where $\beta$ is the weight for the attractive t-force. Following the suggestion of ForceAtlas2~\cite{jacomy2014forceatlas2}, we use a
a linear attractive spring force, and an attractive t-force with $\varphi=1$.

%Here, we set $p$ to 1 for simplifying the spring force.
Combining both forces on the $i$-th node yields the following resultant force:
\begin{footnotesize}
    \begin{align}
        F(i)= & \sum_{i \neq j} -\frac{||\mathbf{x}_{i}-\mathbf{x}_{j}||}{(1+||\mathbf{x}_{i}-\mathbf{x}_{j}||^2)^{\gamma}}
        \frac{\mathbf{x}_{i}-\mathbf{x}_{j}}{||\mathbf{x}_{i}-\mathbf{x}_{j}||} +  \nonumber                                \\
              & \alpha\sum_{i \leftrightarrow j} \left( ||\mathbf{x}_{i} - \mathbf{x}_{j}|| +
        \frac{\beta||\mathbf{x}_{i}-\mathbf{x}_{j}||}{1+||\mathbf{x}_{i}-\mathbf{x}_{j}||^2}
        \right ) \frac{\mathbf{x}_{i}-\mathbf{x}_{j}}{||\mathbf{x}_{i}-\mathbf{x}_{j}||},
    \end{align}
\end{footnotesize}
where  $\gamma$ is the exponent for the repulsive t-force component, and $\alpha$ is the weight for the attractive force.
This model has the same attractive spring forces as the original model, but enhances it with an attractive short-range t-force and  replaces its repulsive force with a repulsive short-range t-force.
%In doing so, it yields similar overall graph structures but better preserve local neighborhoods.
The attractive short-range t-force has a \od{similar form as the one in tsNET (see the supplemental material), which is weighted by an extra term that is derived from graph theoretical distances. In contrast to tsNET, our attractive short-range t-force is only exerted on given edges, and is combined with long-range spring forces for better maintaining global structures (see Section~\ref{sec:comparelayout}).}

\vspace{2mm}
\noindent\textbf{Parameter Analysis}. \
The parameter $\gamma$ specifies the extent and magnitude of the repulsive t-force that controls the longest distance of neighbors in the layout,  while $\alpha$ and $\beta$ tune the weight of the attractive long-range and short-range t-forces, respectively.
To meet the above three principles, they have to be set in a reasonable way.
However, analyzing the relationship between them for large graphs is hard because each node is influenced by potential forces from many other nodes.
For simplicity, we first investigate their valid ranges from the extreme case with two connected nodes and then provide guidelines for general graphs.

\begin{figure}[tb]
    \centering
    \includegraphics[width=0.9\linewidth]{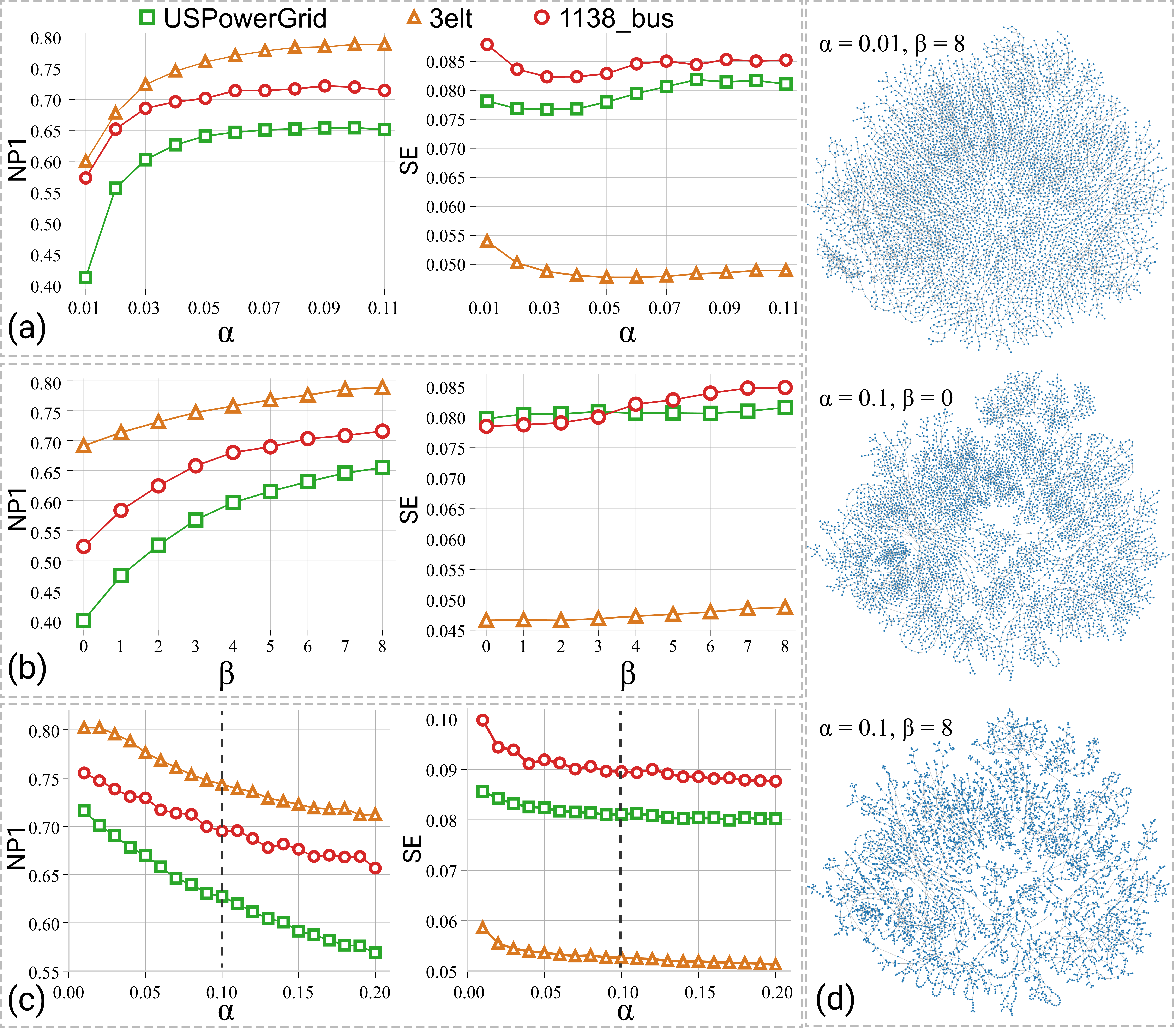}\vspace{-1mm}

    \caption{Influence of $\alpha$ and $\beta$ for graph layouts: (a) For a fixed $\beta = 8$, a small $\alpha$ distributes nodes in the layout of the \textit{USpowerGrid} graph  more evenly;
        (b) For a fixed $\alpha = 0.1$, a large $\beta$ results in more sub-clusters;
        (c) Different combinations of $\alpha$ and $\beta$ for $\alpha\beta = 0.8$ create different 1-ring neighborhood preservation values (higher is better) and  stress errors (lower is better) for the  three given graphs. $\alpha=0.1$ creates a good balance between optimizing both aspects;
        (d) Three layout results of the \textit{USpowerGrid} graph generated by different combinations of $\alpha$ and $\beta$.
    }\vspace{-2mm}
    % \caption{Layout of the USpowerGrid by different parameters. (a) For a fixed $\beta$=8, a small $\alpha$ distributes nodes more evenly. (b) different combinations of $\alpha$ and $\beta$ for a fixed $\alpha\beta$ create different local environments and edge lengths. (c) an exponent of $\gamma$=2 seems to produce good layouts. }
    \label{fig:weightExponents}
\end{figure}

For specifying $\alpha$ and $\beta$, we formally represent \textbf{P2} as the relationship between repulsive and attractive force for two connected nodes $i$ and $j$:
\begin{equation}
    \lim_{|\mathbf{x}_{i}-\mathbf{x}_{j}|\rightarrow 0} \frac{F^a(i,j)}{F^r(i,j)} < 1.
    % \nonumber
    \label{eq:limitf}
\end{equation}
Substituting Eq.~\ref{eq:replusive} and Eq.~\ref{eq:newforce} into Eq.~\ref{eq:limitf} yields:
\begin{equation}
    %    \left\{
    %    \begin{aligned}
    %        \alpha\beta     & <1   \quad \textrm{if}\; p > 1  \\
    %        \alpha(1+\beta) & <1  \quad  \textrm{if}\;  p = 1 \\
    %    \end{aligned}
    %    \right.
    \alpha (1+\beta) < 1 .
    \label{eq:limitweight}
\end{equation}
We can see  that $\alpha$ and $\beta$ cannot be both large, while they cannot be both small so as to balance attractive and repulsive forces.
The smaller $\alpha$, the weaker the long-range attraction and the more likely it is to produce long edges, resulting in the layout with uniformly distributed nodes (see Fig.\ref{fig:weightExponents}(a) which uses $\alpha=0.01$).
Otherwise, long-edges are increased and the layout tends to split into many sub-clusters  (see Fig.\ref{fig:weightExponents}(a) with $\alpha=0.50$).
Increasing $\beta$ results in the stronger the short-range attraction and the more likely it is to group connected nodes together, forming more sub-clusters (see Fig.\ref{fig:weightExponents}(b) with $\beta$ =50). For a fixed $\beta$, a smaller $\alpha$ lets  $\alpha\beta$ decrease and the proportion of repulsive forces increase. In this case the layout tends to distribute nodes more uniformly (see Fig.\ref{fig:weightExponents}(a) which uses $\alpha=0.01$). We can see that there are  similar trends in Figs.\ref{fig:weightExponents}(a,b).

For a fixed $\alpha\beta$, decreasing $\alpha$ increases the proportion of short-range attractive and repulsive forces relative to the overall forces, resulting in a better preservation of  graph neighborhoods (see the \od{amount}
%degree
of the 1-ring neighborhood preservation in the left of Fig.\ref{fig:weightExponents}(b)). Yet, distance preservation will be worse (see stress error in Fig.\ref{fig:weightExponents}(b)) because smaller long-range attractive forces lead to many long edges.
%In all, we summarize that %we find that a configuration of $\alpha$ = 0.1 and $\beta$ = 8 works well in our experiments and preserves local and global structures well.% \odc{(If this is the best configuration we have to show this parameter combination in Figure \ref{fig:weightExponents})}.

\begin{figure}[t]
    \centering
    \includegraphics[width=0.98\linewidth]{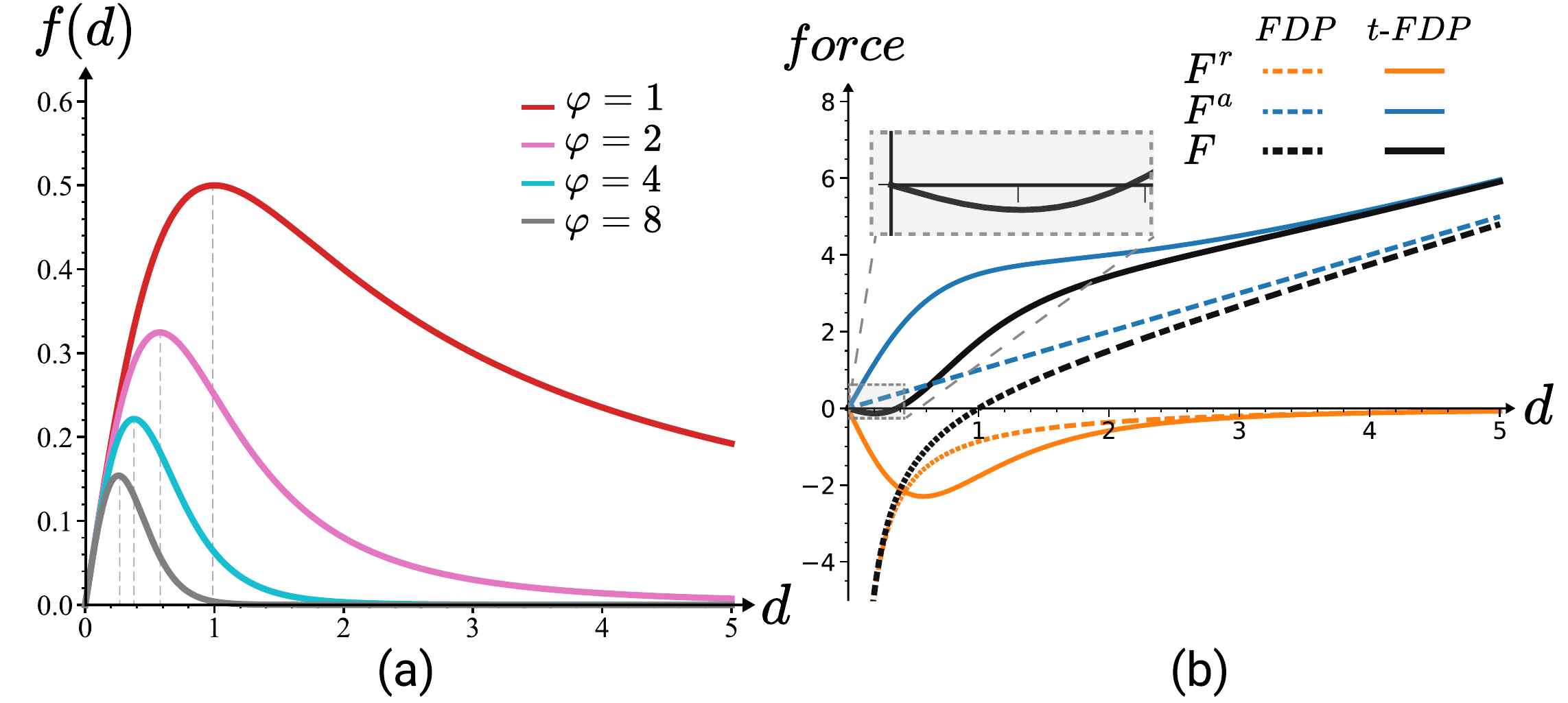}\vspace{-4mm}
    \caption{(a) Function graphs of the t-force for different $\varphi$; (b) comparison of repulsive (orange), attractive (blue) and resultant (black) forces exerted on two connected nodes between FDP and t-FDP shown as the dotted and solid lines, respectively. The major difference are the short-range forces, which are bounded in t-FDP.}
    \label{fig:functiongraphs}\vspace{-5mm}
\end{figure}

For specifying $\gamma$, we formally represent \textbf{P3} which requires the attractive force between two connected nodes to be larger than the repulsive force for forming reliable clusters except when nodes approach each other. Hence, $F^a$ and $F^r$ satisfy the following condition:
\begin{equation}
    F^a(i,j)>F^r(i,j) \quad\quad \textrm{if} \quad \|\mathbf{x}_i -\mathbf{x}_j\|>\epsilon \nonumber
\end{equation}
where $\epsilon$ is a small value but larger than zero.  Since the t-force is a short-range force, the attractive force is certainly larger at long ranges than the repulsive force  because of its spring force component. However, for short distances this condition might not hold.
As shown in Fig.~\ref{fig:functiongraphs}(a), the larger the exponent $\gamma$, the smaller the force magnitude.
To prevent the attractive t-force weighted by $\alpha \beta$ from being smaller than the repulsive force, % (which is less than one determined by Eq.\ref{eq:limitweight})
we require:
%If $\varphi$ is larger than or equal to $\gamma$, the attractive t-force will certainly be smaller than the repulsive force, since it is weighted by $\alpha \beta$, which is smaller than one as determined by Eq.\ref{eq:limitweight}. Thus, we require that:
\begin{equation}
    \gamma > 1  . \label{eq:exponentcondiction}
\end{equation}

\begin{figure}[tb]
    \centering
    \includegraphics[width=0.9\linewidth]{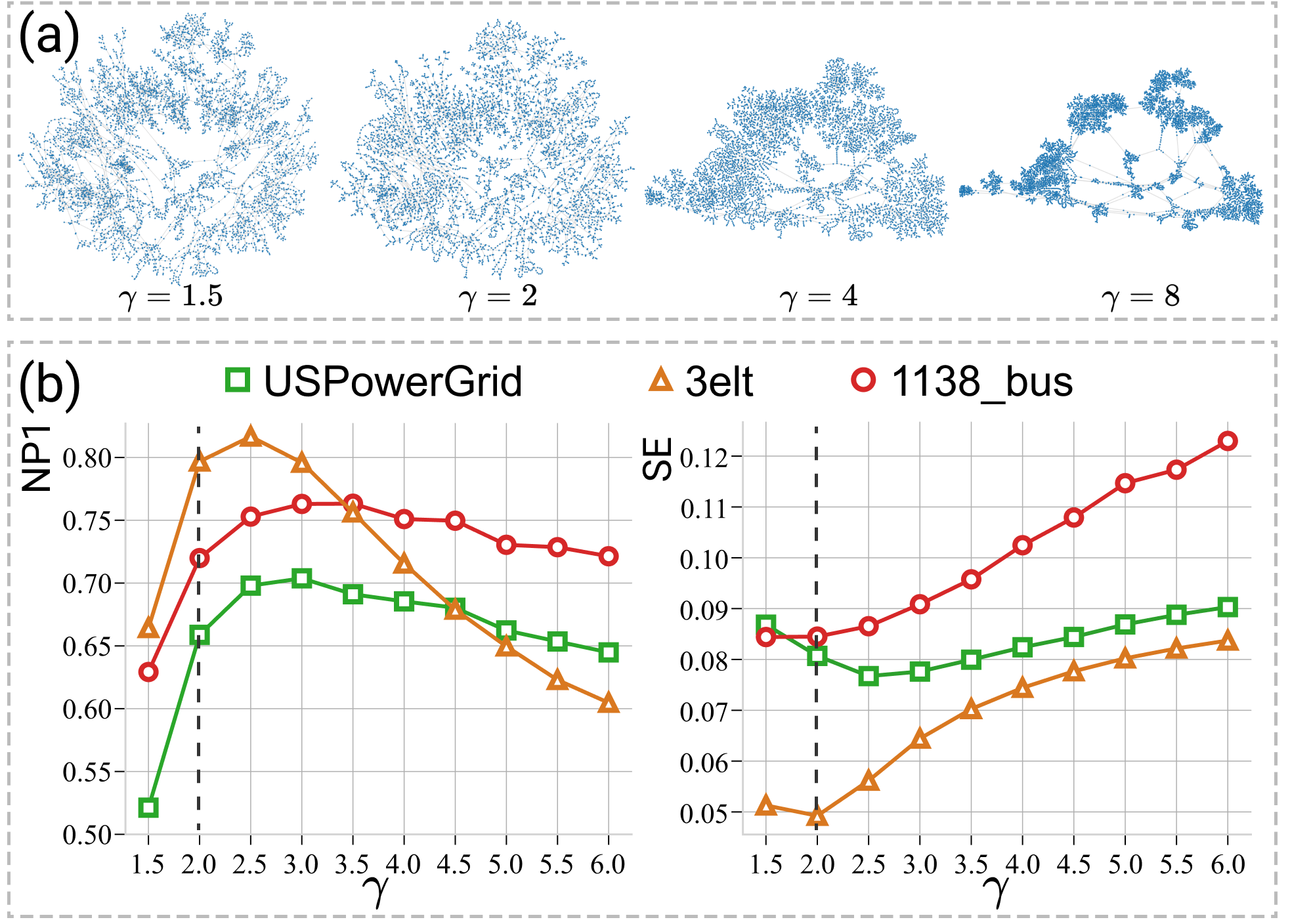}\vspace{-1mm}
    \caption{
        Influence of  $\gamma$ for graph layouts: (a) Layouts of the
        \emph{USpowerGrid} graph for $\gamma=2,4,8$; (b)  1-ring neighborhood preservation(left) and stress error(right) for varying  $\gamma$ for three graphs.  $\gamma=2$ creates a good balance between optimizing both aspects.}\vspace{-2mm}
    \label{fig:exponent}
\end{figure}

As shown in Fig.\ref{fig:functiongraphs}(a), larger exponents $\gamma$ result in repulsive forces with shorter range. In relative terms, the resultant force creates an attraction at closer distances, which leads to better representing local cluster structures (see Fig.\ref{fig:exponent}(a)) but does not efficiently preserve neighborhoods and distances (Fig.\ref{fig:exponent}(b)). %Through practical experiments, we find  $\gamma$ = 2 is a good compromise for most graphs

% %\subsection{Parameter Analysis}
% Following the default setting of existing FDP methods, we use the linear spring forces by setting $p$ to 1. Fig.~\ref{Uspower} show the layout of USpowerGrid by different parameters.

% \noindent\textbf{Exponents.}
% Since  \textbf{P2} requires the repulsion force to be as large as the attraction force, we therefore set the short-range attraction exponent $\varphi$ to 1.

Overall, $\gamma$ specifies the extent and magnitude of the repulsive t-force that controls the longest distance of neighbors in the layout,  while $\alpha$ and $\beta$ tune the weights of the attractive long-range and short-range t-forces, respectively.
In our experiments,  we find that a configuration of $\alpha$ = 0.1, $\beta$ = 8 and $\gamma$ = 2 works well and preserves local as well as global structures. % \odc{(If this is the best configuration we have to show this parameter combination in Figure \ref{fig:weightExponents})}.
Fig.~\ref{fig:functiongraphs}(b) compares attractive, repulsive, and resultant forces exerted on a node using the standard FDP and our t-FDP model. The repulsive force is larger than the attractive force when the distance between two nodes is smaller than a certain value; otherwise, the attractive force is larger. This is well aligned with the three requirements \textbf{R1-R3}.

\vspace{-2mm}
\subsection{Approximate Calculation of Repulsive Forces}
\label{sec:approx}
%\begin{figure*}[ht]
%    \centering
%    \includegraphics[width=\textwidth]{FFT2}
%    % \includegraphics[width=0.48\linewidth]{tdis.pdf}
%    % \subfigure[]{\includegraphics[width=0.48\linewidth]{repforce_current.pdf}}
%    \caption{An example to explain FFT-accelerated interpolation-based approximation. $y_i$ is mapped to a uniform grid using polynomial interpolation and a Toeplitz matrix was obtained by calculating kernel function values between uniform grids. Toeplitz matrix can be
%        embedded into a circulant matrix of twice its size. FFT can be used to quickly calculate multiplication of circulant matrix. Map resulting coefficients vector onto the grid. The approximate results can be collected by the reverse method of interpolation.}
%    \label{fig:FFT}
%\end{figure*}

Computing repulsive forces in a graph involves the interaction between each node and all other nodes, resulting in a computational effort of $O(n^2)$ (we refer to this as the \emph{exact} method).
As proposed in \cite{van2014accelerating} for t-SNE, the computation can be accelerated by using tree-based approximation methods such as  Barnes-Hut (BH)~\cite{barnes86Hierarchical} to truncate the t-forces. However, this still requires $O(n\log n)$ computations, which is infeasible for large graphs. Random vertex sampling~\cite{gove2019random} runs in $O(n)$ time, but the resulting layouts for large graphs are often worse than the ones generated by the other methods. To address this issue, we introduce an interpolation-based Fast Fourier Transform (ibFFT) algorithm, which was originally designed for accelerating t-SNE~\cite{linderman2019fast}. It yields similar layouts as the BH method, but is about ten times faster for most large graphs.
%, having a nearly linear complexity with respect to $n$.
% OD: I removed this since the other methods also have in part O(n)

To do so, we first represent the repulsive t-force as multiple sums of the weighted kernel functions:

\vspace{-2mm}
\begin{align}
    F^r(i)
     & =  \sum_{j=1,j\neq i}^{n}  \frac{{||\mathbf{x}_{i}-\mathbf{x}_{j}||}}{(1+||\mathbf{x}_{i}-\mathbf{x}_{j}||^2)^{\gamma}} \frac{\mathbf{x}_{i} - \mathbf{x}_{j}}{{||\mathbf{x}_{i}-\mathbf{x}_{j}||}}                             & \nonumber \\
     & = \sum_{j=1,j\neq i}^{n}  \frac{\mathbf{x}_{i} - \mathbf{x}_{j}}{(1+||\mathbf{x}_{i}-\mathbf{x}_{j}||^2)^{\gamma}}                                                                                                              & \nonumber \\
     & =\mathbf{x}_{i}\sum_{j=1}^{n} \mathbf{K}(\mathbf{x}_{i},\mathbf{x}_{j}) -  \sum_{j=1}^{n} \mathbf{K}(\mathbf{x}_{i},\mathbf{x}_{j}) \mathbf{x}_{j}                                                             \label{eq:repfK}
\end{align}
% \mathbf{K}
% \mathbf{K}
where the kernel $\mathbf{K}(\mathbf{x}_{i},\mathbf{x}_{j})$ is:
\vspace{-2mm}
\begin{align}
    \mathbf{K}(\mathbf{x}_{i},\mathbf{x}_{j}) = \frac{1}{(1+||\mathbf{x}_{i}-\mathbf{x}_{j}||^2)^{\gamma}}.
\end{align}
Since $\mathbf{x}_{j}$ is a 2D point,  each dimension can be calculated independently:\vspace{-2mm}
\begin{align}
    F^r(i)_{(1)} & =  {\mathbf{x}_{i}}_{(1)}  \sum_{j=1}^{n} \mathbf{K}( \mathbf{x}_{i},\mathbf{x}_{j}) -  \sum_{j=1}^{n} \mathbf{K}(\mathbf{x}_{i},\mathbf{x}_{j}) {\mathbf{x}_{j}}_{(1)} \label{eq:Fr1}   \\
    F^r(i)_{(2)} & =  {\mathbf{x}_{i}}_{(2)}  \sum_{j=1}^{n} \mathbf{K}( \mathbf{x}_{i},\mathbf{x}_{j})  - \sum_{j=1}^{n} \mathbf{K}(\mathbf{x}_{i},\mathbf{x}_{j}) {\mathbf{x}_{j}}_{(2)},  \label{eq:Fr2}
\end{align}
where ${\mathbf{x}_{j}}_{(1)}$  and ${\mathbf{x}_{j}}_{(2)}$ are  the x-coordinate  and y-coordinate of $\mathbf{x}_{j}$, respectively.
Hence, $F^r(i)$ is determined by three sums of products between the kernel $\mathbf{K}(\mathbf{x}_{i},\mathbf{x}_{j})$ and the weight $v_j$:
\vspace{-2mm}
\begin{align}
    \psi(\mathbf{x}_{i}) = \sum_{j=1}^{n} \mathbf{K}(\mathbf{x}_{i},\mathbf{x}_{j})v_j.
    \label{eq:kernelproduct}
\end{align}
where $v_j$ is 1, ${\mathbf{x}_{j}}_{(1)}$, or ${\mathbf{x}_{j}}_{(2)}$.
For example, $F^r(i)_{(1)}$ is computed by two sums of products in Eq.~\ref{eq:Fr1} with the weights $v_j$ being 1 and ${\mathbf{x}_{j}}_{(1)}$.
Using the naive way to compute such a sum takes $n^2$ time, which is prohibitive for a large number of nodes $n$.

%\odc{Since most of the following (in other form) is already given in the Nature Methods paper, we could move it to the appendix if we need space}
Alternatively, Linderman et al.~\cite{linderman2019fast} exploited the low-rank nature of the kernel $\mathbf{K}$ to approximate Eq.~\ref{eq:kernelproduct} with a small number ($k\times k$) of equi-spaced 2D grid nodes.
This is done in three steps:
\begin{itemize}
    \setlength\itemsep{1mm}
    \item  projecting all data points $\mathbf{x}_i$ onto the grid by using Lagrange polynomials with a time complexity $O(k^2n)$;
    \item computing the interaction of the grid nodes, which can be accelerated by FFT with a complexity $O(2 k^2\log k)$; and
    \item back-projecting the interaction of all grid nodes to the original points with time complexity $O(k^2n)$.
\end{itemize}
We can see that the overall time complexity is $O(k^2n)$ where $k$ is a small number.
%In summary, the repulsive force can be computed by first interpolating nodes onto an equi-spaced grid with Lagrange polynomials and subsequently applying the FFT to perform the transformation.

\begin{figure}[htb]
    \centering
    \includegraphics[width=0.9\linewidth]{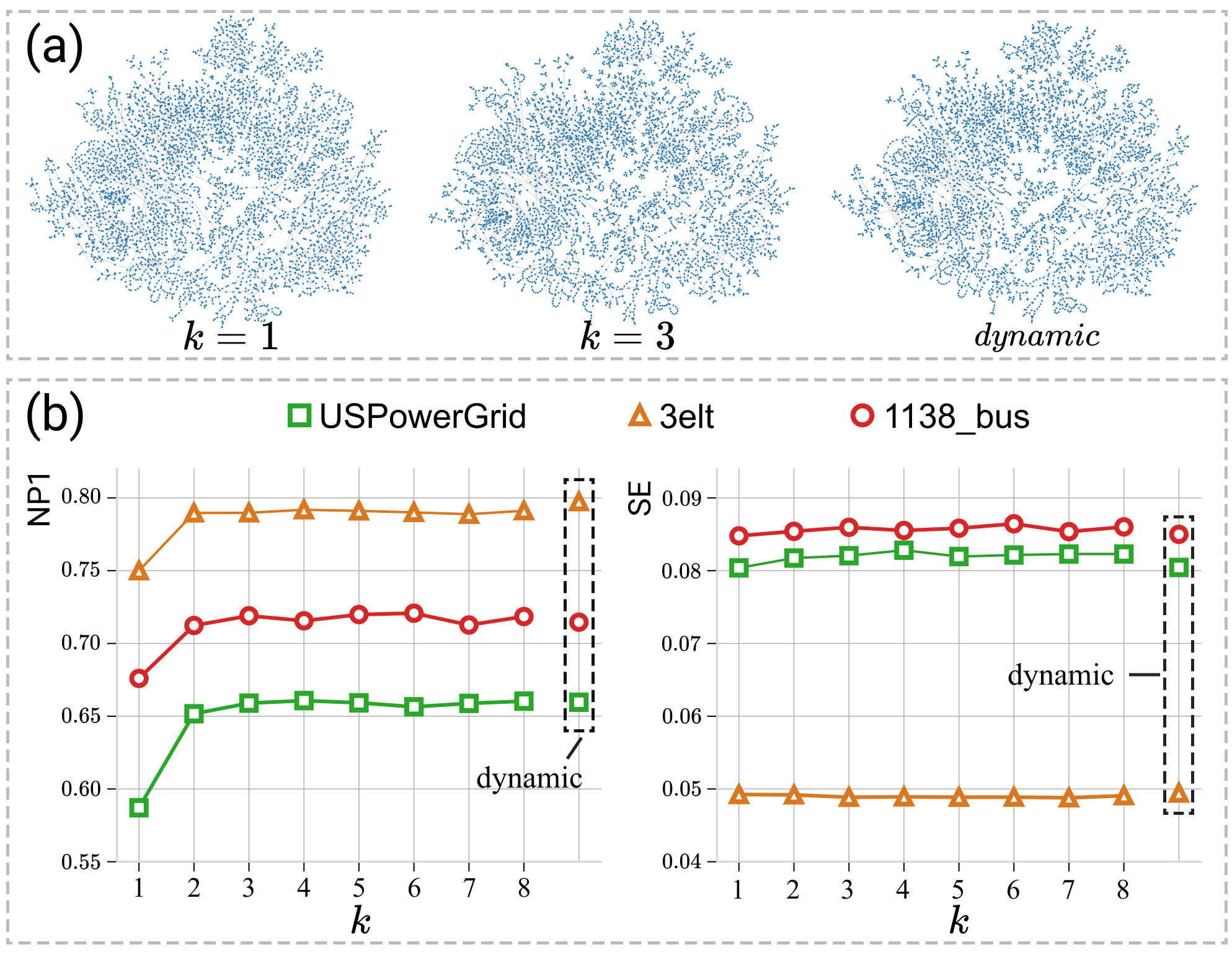}\vspace{-1mm}
    \caption{
        Influence of different  $k$ values on the layout. (a) Layouts of the \textit{UspowerGrid} graph generated by using $k=1$, $k=3$, and our dynamic strategy; (b) 1-ring neighborhood preservation(left) and stress error(right) vary with  $k$ on three graphs. The values of our dynamic strategy are in the dotted boxes.}\vspace{-2mm}
    \label{fig:UspowerFFTK}
\end{figure}

However, the above polynomial interpolation scheme suffers from the Runge phenomenon~\cite{epperson1987runge} which reduces the accuracy as $k$ increases. To overcome this issue,  Linderman et al.~\cite{linderman2019fast}
proposed to %split the interval of all data points into $N_\textrm{int}$  sub-intervals and use $k$ interpolation points within each interval to interpolate each $\mathbf{x}_i$.
%In doing so, the complexity  becomes $O(2kn + (N_\textrm{int} k)\log (N_\textrm{int}  k))$.
%For solving 2D layout,  this strategy has been extended to 2D space by first
divide the interval of the whole layout space $[\mathbf{x}_{\min}, \mathbf{x}_{\max}]\times[\mathbf{x}_{\min}, \mathbf{x}_{\max}]$  into a collection of $N_\textrm{int}\times N_\textrm{int}$ equal sized intervals and then apply polynomial interpolation in each interval with $k\times k$ equi-spaced nodes. Specifically, the computation is done by first projecting each data point $\mathbf{x}_i$ into the corresponding grid square, then computing the interaction of the $k\times k$ equi-spaced grid nodes within each grid square; and finally back-projecting the interaction of all grid nodes within each grid square to the original points.
The first and last steps still requires $O(k^2n)$ operations, while the second one can be done in $O(2(N_\textrm{int}k)^2 log(N_\textrm{int}k))$ operations with the FFT acceleration.
Hence, the total computational complexity becomes $O(nk^2+(N_\textrm{int}k)^2 log(N_\textrm{int}k))$.
%into a collection of $N_\textrm{int}\times N_\textrm{int}$ equal sized squares, we apply
%For more detail, please refer to the supplemental material.

\vspace{2mm}
\noindent\textbf{Choices of $k$}.
Linderman et al.~\cite{linderman2019fast} suggested to set $k=3$ and $N_\textrm{int} = \max (50, [y_\textrm{max}- y_\textrm{min}])$ for t-SNE when visualizing high dimensional data.
We found that the layouts generated by  using $k=3$  are similar to the ones
produced by using $k=1$ for displaying global structures (see Fig.~\ref{fig:UspowerFFTK}(a)) but  better in neighborhood preservation (left of Fig.~\ref{fig:UspowerFFTK}(b)).
However, the computational complexity quadratically increases with $k$.

To find a good trade-off, we suggest to vary the values of $k={1,2,3}$  during the optimization: we start with $k=1$ for 90\% of the iterations to obtain a reasonable layout and then use $k=2$ for 5\% of the iterations  and finally $k=3$ for the remaining iterations. An example is shown in the right of Fig.~\ref{fig:UspowerFFTK}(a), which has almost the same structure as the one generated by directly using $k=3$ (see the middle in Fig.~\ref{fig:UspowerFFTK}(a)). We tested multiple datasets and found that such dynamic changes in $k$ values generates high-quality layouts (see values for the dynamic strategy in Fig.~\ref{fig:UspowerFFTK}(b)) with the least computation time. The computation time of the dynamic strategy is close to that of the setting $k=1$.

Based on such choices of $k$, our t-FDP can run in O(n) time, especially for large graphs with $N_\textrm{int}$ being much smaller than $n$.

\section{Evaluation}\label{sec:evaluate}
Following UMAP~\cite{mcinnes2018umap}, we implemented t-FDP in \emph{Python3} and used the \emph{numba} library~\cite{lam2015numba} compiler to translate the Python code to fast machine code with a similar speed as plain C code. To accelerate it using the GPU, we used the open-source matrix library \emph{cupy} to speed up  matrix computations with NVIDIA CUDA. To accommodate  a wide range of users, we further provide a Javascript implementation as a drop-in force for the ``d3-force'' library, which allows for accelerating it with WebGL if a GPU is available. In the following, we refer to the interpolation-based Fast Fourier Transform (ibFFT) implementation of t-FDP by \textit{ibFFT}.

To demonstrate its effectiveness, we evaluate t-FDP in two ways. First, we validate our FFT-based approximation by comparing it with the exact method and other approximations. Second, we compare it with a few state-of-the-art layout methods on a set of synthetic and real-world graphs of various sizes. All results were measured on a desktop machine with Intel i7-8700 processor, 32 GB CPU memory, and NVIDIA GeForce RTX 2080 GPU (8GB GPU memory).

\vspace{1.5mm}
\noindent\textbf{Datasets.}
For a comprehensive evaluation, we collected 50 graphs with varying numbers of nodes and edges (see Table 1 in the supplemental material). Most of them are selected from the Florida Collection~\cite{davis2011university}, the SNAP network collection~\cite{snapnets} and the collections used by tsNET~\cite{kruiger2017graph} and DRGraph~\cite{zhu2020drgraph}.
As for FDP methods, our t-FDP model can inherently handle multi-component graphs and hence we include 7 graphs with more than one component.
In addition, we synthesized three graphs with various cluster structures by using the graph-tool python library~\cite{peixoto2014graph}.

\vspace{1.5mm}
\noindent\textbf{Methods.}
The comparison includes ten graph layout methods:  force-directed placement by Fruchterman-Reingold (FR)~\cite{fruchterman1991graph},
force-directed layout by random vertex sampling (FR-RVS)\cite{gove2019random},
SFDP~\cite{hu2005efficient}, ForceAtlas2 (FA2)~\cite{jacomy2014forceatlas2} , Linlog~\cite{noack2007energy},  PivotMDS (PMDS)~\cite{brandes2006eigensolver}, stress model (SM)~\cite{Khoury12Drawing}, the maxent-stress model (Maxent)~\cite{gansner2012maxent}, tsNET~\cite{kruiger2017graph} and DRGraph~\cite{zhu2020drgraph}. Including our methods, we can categorize all methods into three groups: force-based methods (FR, SFDP, FA2, LinLog, and t-FDP), stress-based methods (PMDS, SM, Maxent) and neighboring embedding methods (tsNET and DRGraph). %Four force-based methods (FR, SFDP, FA2, LinLog) set the exponents $(p,q) $of the power functions defined in Eq.~\ref{eq:powerattractive} and Eq.~\ref{eq:powerreplusive} are (2,1), (2,2), (1,1) and (0,1), respectively.
We use the public C++ libraries OGDF~\cite{chimani2013open} and GraphViz~\cite{gansner2009drawing} for performing FR, SM, and SFDP while taking the implementations of FA2, Linlog, Maxent, tsNET and
DRGraph from the authors of the original papers. We implemented PMDS by ourselves instead of using the method provided by OGDF, since it was not stable and did not allow to efficiently handle large data. tsNET was accelerated  with a GPU-based t-SNE implementation~\cite{chan2018t}.
For FR-RVS, the public Javascript implementation cannot handle large graphs and we re-implemented it with Python and used the Numba~\cite{lam2015numba} library for acceleration.
We ran each method with  the default parameters provided by the original paper or implementation. For example, we set the perplexity of tsNET to 40 and choose the first-order nearest neighborhood for DRGraph.

For fair comparisons, we use the same PMDS layout as initialization for all methods except SFDP and DRGraph, which are initialized by a multi-level scheme.
Since SFDP and DRGraph are inherently random, we assess the average quality at runtime of each method on one graph by calculating the average over 5 runs.
We provide the results of a random initialization in the supplemental material, they show that most of the methods can be significantly improved by using the PMDS layout as initialization, \yh{which is consistent with the previous finding in t-SNE and UMAP~\cite{kobak2021initialization}.}
%For each dataset, we run all methods five times with different initializations and report the average result.

\begin{figure}[!t]
    \centering
    \includegraphics[width=1.0\linewidth]{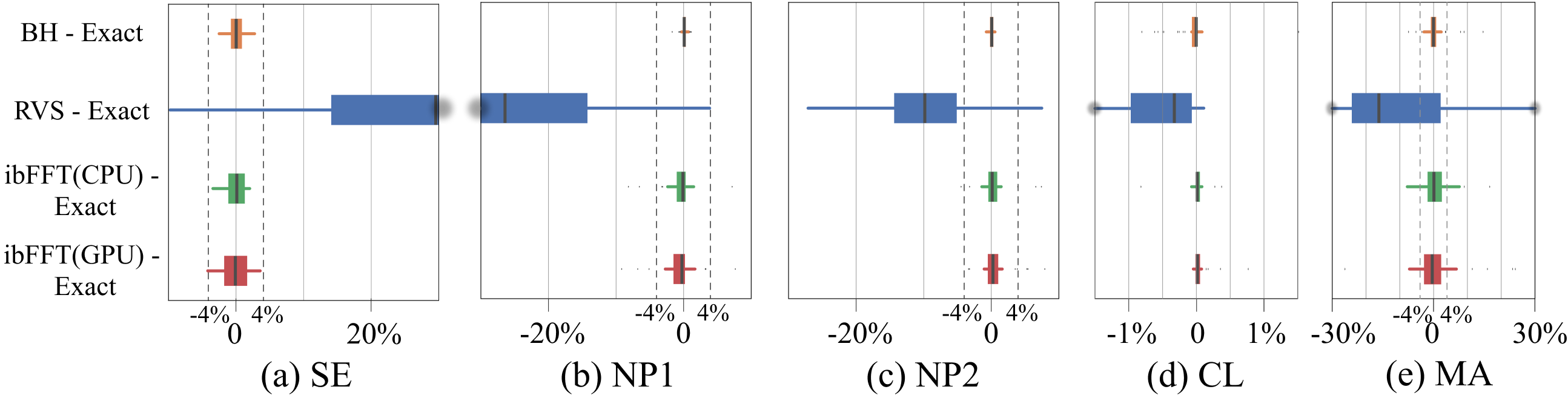}
    \caption{Five relative error metrics (abbreviations see text) for  layouts generated by comparing four approximation methods with the exact method of the t-FDP model}. \vspace{-4mm}
    \label{fig:approxQuality}\vspace{-2mm}
\end{figure}

\vspace{1.5mm}
\noindent\textbf{Metrics.}
Following the evaluations of tsNET~\cite{kruiger2017graph}  and DRGraph~\cite{zhu2020drgraph}, we evaluate graph layouts with four metrics: normalized stress error, neighborhood preservation degree, crosslessness, and minimum angle. The first two metrics characterize the graph structure in preserving distances and neighborhoods, while the latter two reflect graph readability in terms of edge crossing and minimum angle.
\begin{itemize}
    \item \emph{Normalized stress error} (SE) \cite{gansner2012maxent}
          To assess the distance-preservation of a layout, we use the normalized stress error defined as:
          \begin{equation}
              SE = \frac{2}{n(n-1)} \sum_{i < j}\frac{(\bar{s} \|\mathbf{x}_i - \mathbf{x}_j\| - d_{ij})^2}{{d_{ij}}^2}, \nonumber
          \end{equation}
          where $d_{ij}$ is the shortest path distance between the nodes $i$ and $j$ and $\bar{s}$ is a scaling factor to uniformly scale the layout.
          When the nodes $i$ and $j$  are unreachable in the graphs with multiple components, \yh{we follow DRGraph~\cite{zhu2020drgraph} that} defines $d_{ij}$ as infinity and only computes stress errors for the node pairs with finite distances.
          To make a fair comparison, the optimal scaling factor $\bar{s}$ is obtained by minimizing the normalized stress error. For more details, please refer to Gansner et al.~\cite{gansner2012maxent}.%by calculating its derivative with respect to $\bar{s}$:

          % Since SE uses all pairwise distances, it is also used to somewhat represent the global structure preserving.
    \item \emph{Neighborhood Preservation Degree (NP).} \
          To measure layout quality in preserving neighborhood,  the neighborhood preservation degree is defined as the Jaccard similarity between the input graph and $k_i$-nearest neighborhood graph defined in the layout:
          \begin{equation}
              NP = \frac{1}{n} \sum_{i} \frac{N_G(i, r) \cap N_L(\mathbf{x}_i,k_i) }{N_G(i, r) \cup N_L(\mathbf{x}_i,k_i)}, \nonumber
          \end{equation}
          where $N_G(i, r)$ are $r$-ring neighbors of node $i$ in graph space, $k_i$ is  $|N_G(i, r)|$,
          and $N_L(\mathbf{x}_i,k_i)$ is a set of nodes corresponding to the $k_i$-nearest-neighbors of $\mathbf{x}_i$ in the layout space.
          In our study, we compute the neighborhood preservation degrees with 1-ring and 2-ring graph neighborhoods and denote them as \emph{NP1} and \emph{NP2}, respectively.

    \item \emph{Crosslessness (CL).} \
          To indicate the degree of edge crossing,  the crosslessness metric~\cite{purchase2002metrics} is defined as a normalized value of the number of edge crossings:
          \begin{equation}
              CL =  \left\{
              \begin{aligned}
                   & 1 - \sqrt{c/c_{max}}, & \ \textrm{if}\ c_{max} > 0 & \\
                   & 1 ,                   & \textrm{otherwise}  .      &
              \end{aligned} \nonumber
              \right.
          \end{equation}
          Here, $c$ is the number of edge crossings, and $c_{max}$ is the theoretical upper bound of this  number in each graph. A larger value indicates a better layout with edge crossing.
          %\begin{equation}

    \item \emph{Minimum Angle (MA).}
          The minimum angle metric~\cite{purchase2002metrics} computes the mean deviation between the actual minimum angle and the ideal minimum angle between edges:
          \begin{equation}
              MA = 1 - \frac{1}{n} \sum_{i}^{n} \frac{|\theta(i) - \theta_{min}(i)| }{\theta(i)}, \quad \theta(i) = \frac{2\pi}{\textrm{degree}(i)} \nonumber
          \end{equation}
          where  $\theta_{min}(i)$ is actual minimum angle at the node $i$. $MA$ is in the range [0,1] and reaches the maximum when all the nodes have equal angles between all incident edges.
\end{itemize}

For comparing different methods,  we calculated the relative values for the above-given metrics
computed from layouts generated by a source $s$ and a target method $t$.
Taking the $SE$ metric as an example, the relative error between two layouts $\mathbf{X}_s$ and  $\mathbf{X}_t$ of one graph is:
\begin{equation}
    \overline{SE}=\frac{SE(\mathbf{X}_s) - SE(\mathbf{X}_t)}  {SE(\mathbf{X}_t)},  \nonumber
\end{equation}
for any $SE(\mathbf{X}_t)$ larger than zero. %which will be a value between -1 and 1\fhz{The range of values of relative error is between -1 and infinity, I think we maybe needn't describe the range of values. Although it will generally be between -1 and 1, it is incorrect}.
Since a smaller stress error indicates better distance preservation, a negative relative error reflects that the source method performs better. For the other metrics, positive relative errors indicate that the source method performs better.

\begin{figure}[!t]
    \centering
    \includegraphics[width=1.0\linewidth]{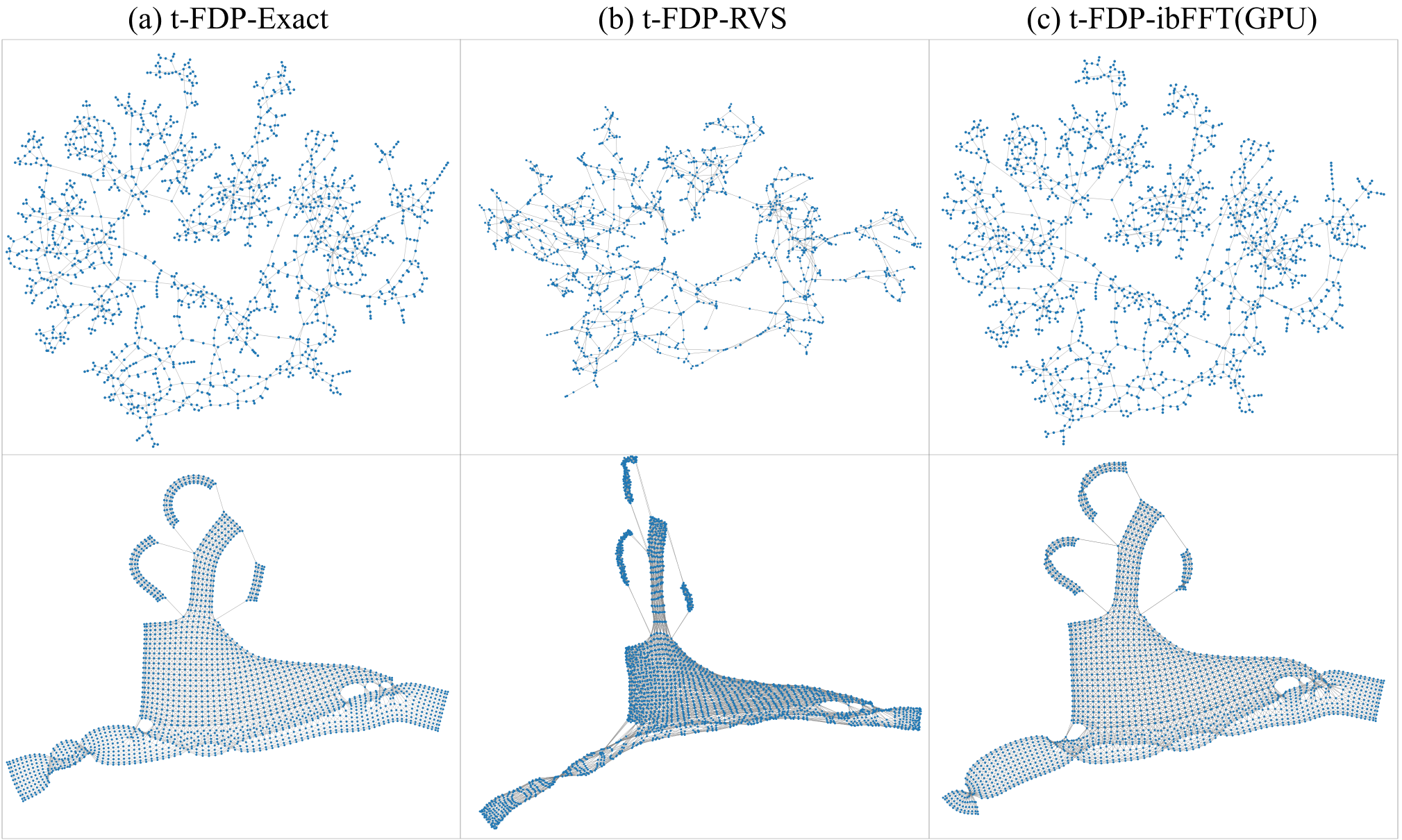}
    \caption{Visual results for \textit{bcspwr07} and \textit{walshaw-1}  generated by  the exact t-FDP model (a) and two approximation methods: t-FDP-RVS (b) and t-FDP-ibFFT(GPU) (c). }\vspace{-2mm}
    \label{fig:approxVis}
\end{figure}

\subsection{Comparison of Approximation Methods}
The goal of this experiment was to measure the efficiency of the CPU and GPU versions of our ibFFT algorithm. It is done by comparing four approximation methods (BH~\cite{barnes86Hierarchical}, RVS~\cite{gove2019random}, and the CPU and GPU versions of our ibFFT algorithm) with the exact method in terms of the quality statistics, convergence  rate and runtime performance. For simplicity, we refer five methods as \emph{t-FDP-BH}, \emph{t-FDP-RVS},
\emph{t-FDP-ibFFT(CPU)}, \emph{t-FDP-ibFFT(GPU)} and t-FDP-exact, respectively.
% For measuring the layout quality, we do not include the graphs with the nodes larger than 100K, since the naive method takes more than three hours to compute the layout for them.
%To measure the accuracy, we calculated the relative values for the above-given measures
%computed from layouts generated by a source $s$ and a target method $t$.
%Taking the $SE$ measure as an example, the relative error between two layouts $\mathbf{X}_s$ and  $\mathbf{X}_t$ of one graph is:
%\begin{equation}
%    \overline{SE}=\frac{SE(\mathbf{X}_s) - SE(\mathbf{X}_t)}  {SE(\mathbf{X}_t)},  \nonumber
%\end{equation}
%for any $SE(\mathbf{X}_t)$ larger than zero. %which will be a value between -1 and 1\fhz{The range of values of relative error is between -1 and infinity, I think we maybe needn't describe the range of values. Although it will generally be between -1 and 1, it is incorrect}.
%Since a smaller stress error indicates better distance preservation, a negative relative error reflects that the source method performs better. For the other measures, positive relative errors indicate that the source method performs better.

\vspace{1.5mm}
\noindent\textbf{Quality Statistics.} \
The boxplots in Fig.~\ref{fig:approxQuality} summarize the relative values of five layout quality metrics between four pairs of counterparts.

Except the comparison to t-FDP-RVS, we can see that the relative values of all metrics are in the range [-4\%, 4\%] with a mean value very close to zero.
Compared to t-FDP-exact,  t-FDP-BH yields the smallest interquartile ranges, followed by two versions of our methods, while t-FDP-RVS performs the worst. Our methods result in slightly more negative relative values for SE and positive relative values for NP2 and CL,  indicating that their resulting layouts are similar or even slightly better than the ones of the exact method for some graphs. In contrast, t-FDP-RVS results relative values for SE that are around 30\% and the ones for NP1 are smaller than -20\%, indicating that the resulting layouts are considerably  worse than the ones of the exact method.
We speculate such lower NP1 values are due to our short-range repulsive forces, where a subset of repulsive forces randomly sampled by RVS might not be enough to repel non-neighborhood nodes.
Note that the difference between the two versions of t-FDP-ibFFT is caused by floating point finite precision on GPUs.

Fig.~\ref{fig:approxVis} shows the results generated by applying the t-FDP-exact, t-FDP-RVS and t-FDP-ibFFT (GPU) on two graphs, where t-FDP-ibFFT (GPU) produces almost the same results as the exact method and performs better than  t-FDP-RVS in revealing the tree- and grid-like structures of the two graphs.
We speculate that computing our bounded short-range repulsive forces using randomly sampled nodes might not be enough to characterize the graph structures.

\begin{figure}[!t]
    \centering
    \includegraphics[width=0.98\linewidth]{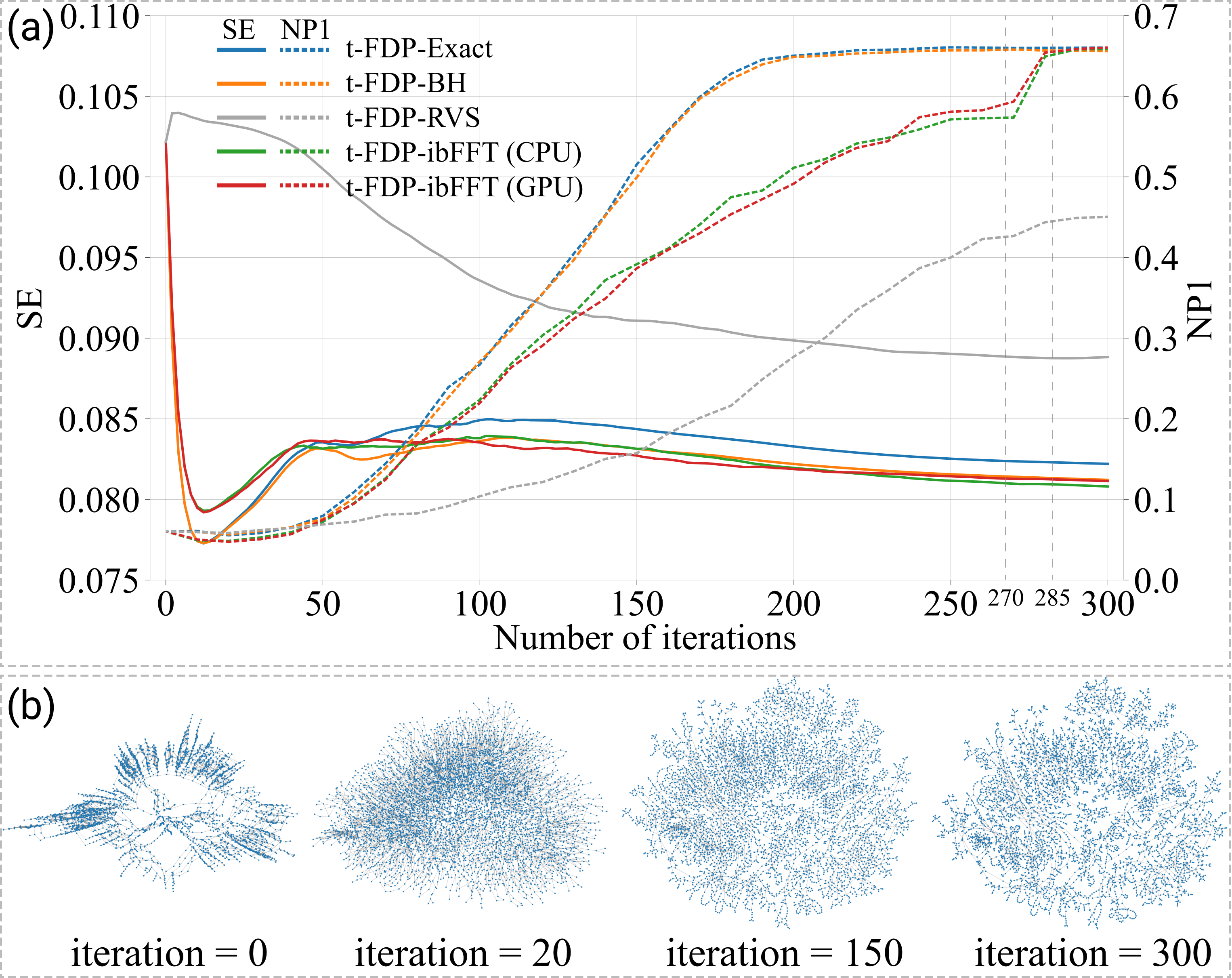}
    \caption{(a) Convergence of  stress error (SE, solid lines) and neighborhood preservation degree (NP1, dashed lines)  for applying the exact method and four different approximation methods of the t-FDP model to the \textit{UspowerGrid} graph; (b) the layout results generated at four different iterations.  }\vspace{-2mm}
    \label{fig:convergence}\vspace{-2mm}
\end{figure}

\vspace{1.5mm}
\noindent\textbf{Convergence Rate.}\
To further inspect the differences between these methods, we investigate the convergence rate of SE and NP1 on the  USPowerGrid graph  for five different methods.
Since these methods have different runtime per iteration, we explore convergence per iterations (see Fig.~\ref{fig:convergence}) and found that
%As the number of interactions increases, all methods achieve the convergence.
all methods show good convergence with increasing number of iterations. The convergence per unit of time of these methods can be found in the supplemental material.
With regard to SE, t-FDP-BH and t-FDP-ibFFT behave similarly and perform slightly better than t-FDP-exact.
Regarding NP1, two versions of t-FDP-ibFFT perform worse than the others for the first 270 iterations and then quickly almost reach their results. This is due to using only a single interpolation point in the first 90\% iterations and then using two and three interpolation points in the remaining iterations (see Section~\ref{sec:approx}). In all, our methods yield a similar neighborhood preservation but a slightly better distance preservation than the exact method and its BH approximation. We speculate that this is because of the random noise induced by our approximation methods, which might improve the optimization quality~\cite{schneider2007stochastic}. The supplemental material shows all results generated by these methods, which have highly similar structures.
Conversely, t-FDP-RVS behaves quite different from the other methods and yields larger stress errors and lower NP1 values, which is consistent with the result shown in Fig.~\ref{fig:approxQuality}.

\begin{figure}[!t]
    \centering
    \includegraphics[width=0.98\linewidth]{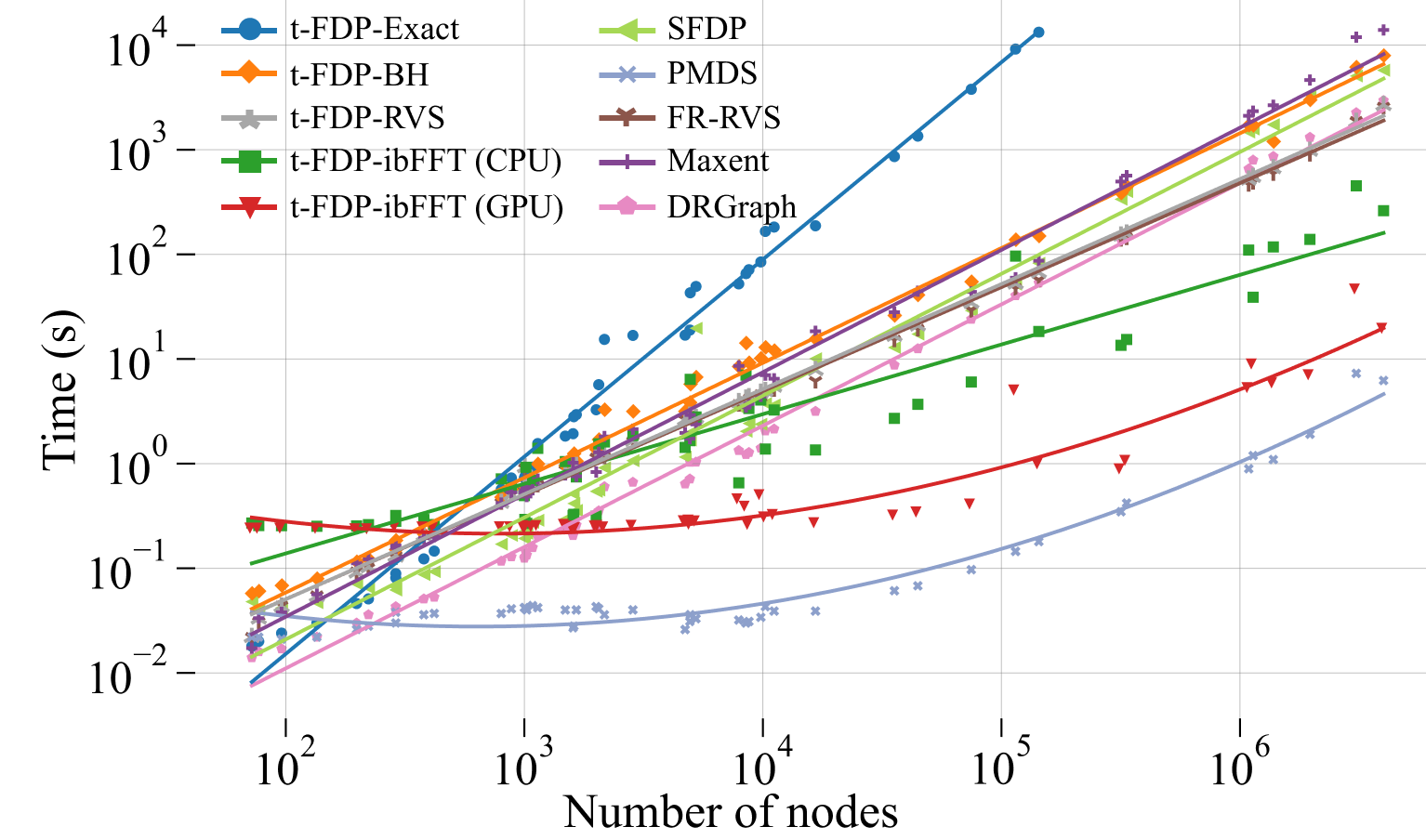}
    \caption{Computation time of five different implementations of our t-FDP model in comparison to five other methods, which can process all datasets.}\vspace{-2mm}
    \label{fig:Performance}\vspace{-2mm}
\end{figure}

\begin{figure*}[!t]
    \centering
    \includegraphics[width=0.96\linewidth]{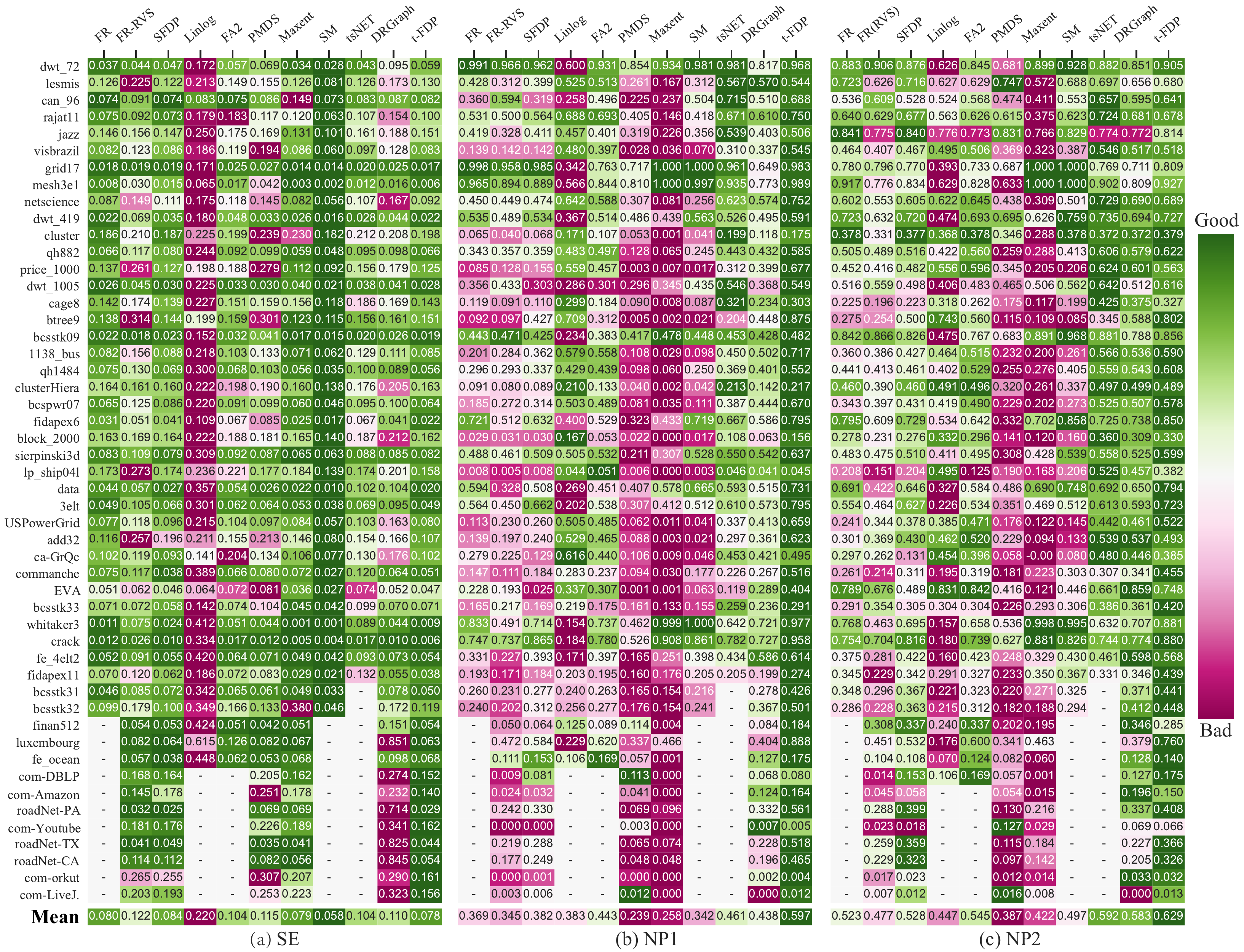}\vspace{-4mm}
    \caption{Heatmaps with a color-blind friendly pink-to-green colormap are used to present the values of SE (a), NP1 (b), and NP2 (c) for layouts generated by ten layout methods on 50 datasets, where the empty cell indicates the graph is too large to be processed by the corresponding layout method.  Each row represents a dataset, and each column a layout method. All rows are colored relatively with regard to best and worst value.
    }\vspace{-4mm}
    \label{fig:heatmap}
\end{figure*}

\begin{figure}[!t]
    \centering
    \includegraphics[width=0.96\linewidth]{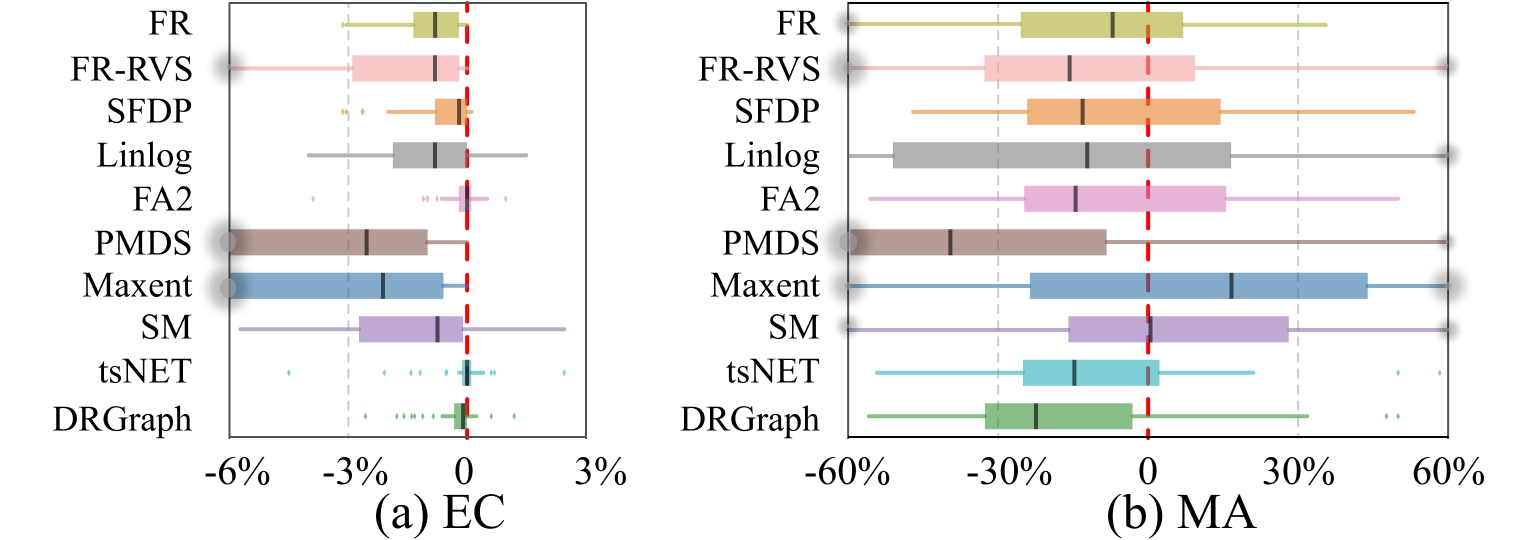} \vspace{-3mm}
    \caption{Boxplots summarizing the values of the two relative error metrics EC and MA over 37 graphs that can be processed by all layout methods. Both errors are relative to t-FDP, large values are better.}
    \label{fig:aQuality}
    \vspace{-5mm}
\end{figure}

\vspace{1.5mm}
\noindent\textbf{Runtime Performance.} \
Regarding runtime performance, the curves in Fig.~\ref{fig:Performance} present the relationship between runtime and number of graph nodes. When the number of nodes is smaller than 500,
both versions of t-FDP-ibFFT take %are an order of magnitude slower than DRGraph and SFDP but
a runtime of 300ms, which is tolerable but slower than t-FDP-exact.
%In contrast, the exact t-FDP has a similar runtime as the other methods.
Beyond this, the benefit of ibFFT is clearly demonstrated.
We can see that t-FDP-ibFFT (GPU) is one order of magnitude faster than t-FDP-ibFFT (CPU)  for large graphs, which is also one order of magnitude faster than the t-FDP-BH. %SFDP, Maxent and DRGraph methods.
In contrast to that, the exact method is the slowest and takes more than 3 hours to solve a graph with 100K nodes, whereas our GPU version can generate the layout in less than 10 seconds for most graphs with millions of nodes.

%While PMDS as a stress model is a very fast method, it cannot compare to the other methods with regard to all error metrics (see below).
% I think we shouls say something here, because PMDS is still the fastest method

Overall, t-FDP-ibFFT (GPU) generates similar layouts as methods like t-FDP-BH, while being much faster for large graphs. Hence, we use this implementation to compare with other layout methods on all datasets.

%%%%%%%%%%%%%%%%%%%%%%%%%%%%%%%%%%%%

\subsection{Comparison of Layout Methods}\label{sec:comparelayout}
Screenshots of the layouts generated by all methods on various graphs with complete scores can
be found in the supplemental material. In the following, we compare the layout methods in terms of layout quality statistics, visual results and runtime performance.

\vspace{1.5mm}
\noindent\textbf{Quality statistics.}
The heatmaps in Fig.~\ref{fig:heatmap} present the SE, NP1, and NP2 values generated by eleven different layout methods for 50 graphs, the other metrics can be found in the supplemental material.
Each row corresponds to a graph, sorted by their number of nodes, and each column corresponds to a graph layout method. Each cell shows a numerical value with the background color encoding the relative metric on the same row and the empty one indicates that the graph is too large to be processed by the corresponding layout method.
We can see that only six methods (FR-RVS, SFDP, PMDS, Maxent, DRGraph and our t-FDP) can handle all graphs. Our t-FDP performs similarly as FR and SFDP in stress errors for most data and performs similarly as tsNET in neighborhood preservation when the number of nodes is less than 500. With the increasing number of nodes, its advantage over the other methods becomes more evident.

Fig.~\ref{fig:heatmap}(a) shows that %t-FDP behaves similarly to FDP  in producing small stress errors for all tested graphs, especially for the large ones.
Maxent and SFDP work well for most graphs but yield large stress errors for some (e.g., \emph{add32}), whereas tsNET and DRGraph generate large stress errors for most graphs and even the worst layout for some examples. We assume that both methods are based on local neighborhoods, while lacking long-range attraction forces for distance preservation.

%\begin{figure}[htb]
%    \centering
%    \includegraphics[width=0.96\linewidth]{tsNETErr.pdf}
%    \caption{\yh{The error bars summarize NP1 and NP2 produced by t-FDP and tsNET with varying perplexity in (a) and the ones produced by t-FDP and DRGraph with varying $k$-order nearest neighbor in (b).}}
%    \vspace{-2mm}
%    \label{fig:diffpara}
%\end{figure}
\begin{figure*}[!t]
    \centering
    \includegraphics[width=0.96\linewidth]{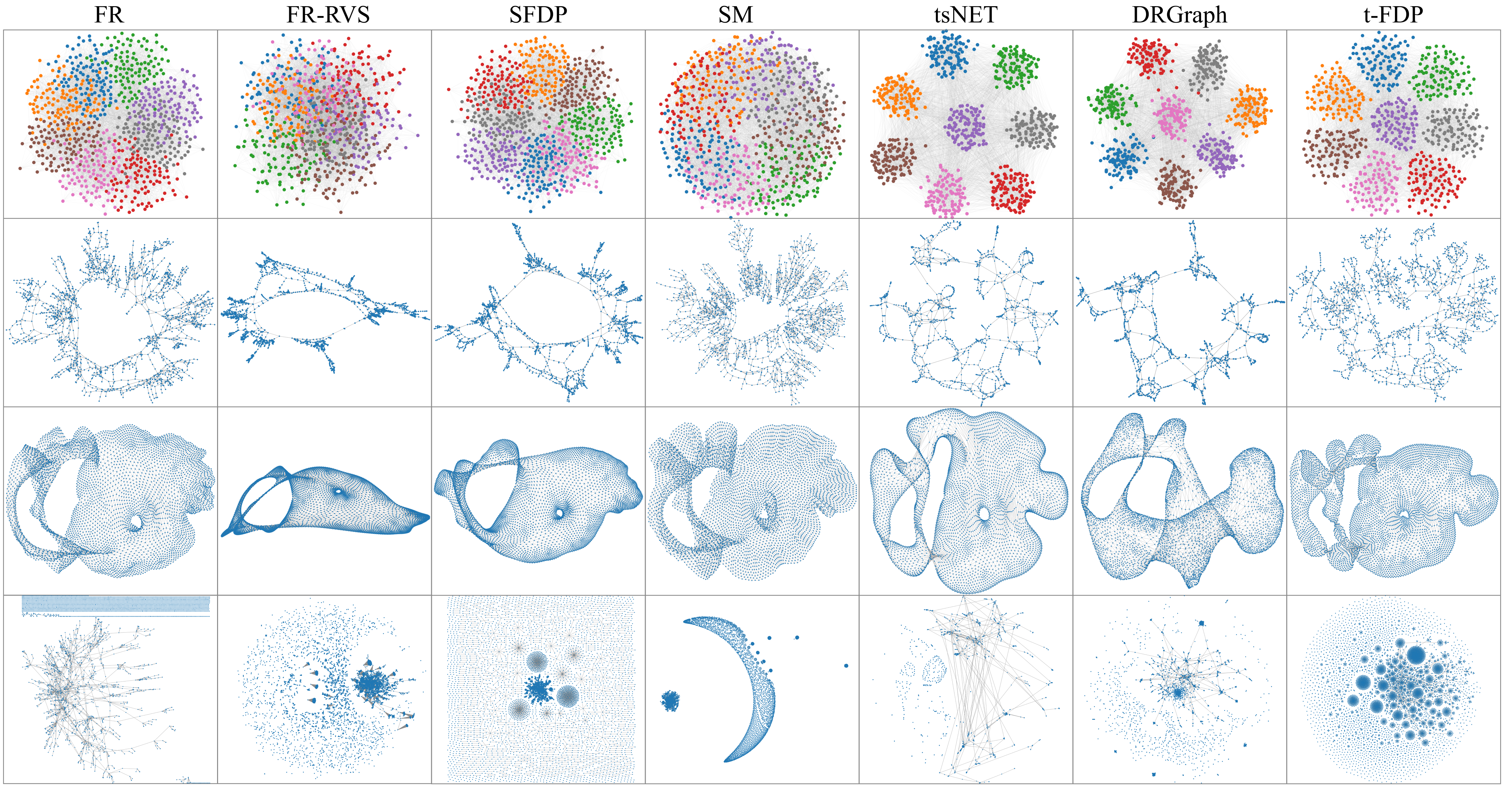}\vspace{-2mm}
    \caption{Layouts by seven methods for the four data sets: \emph{cluster} (top row), \emph{bcspwr07} (second row),  \emph{3elt} (third row), and \emph{eva}(bottom row). t-FDP shows a good ability to highlight clusters, at the same time a good mixture between local and global structures for bcspwr07 and a good unfolding for 3elt. }
    \vspace{-3mm}
    \label{fig:visual}
\end{figure*}

Figs.~\ref{fig:heatmap}(b,c) show that our t-FDP is the best for almost all graphs with regard to NP1  and the best or second-best for NP2 for most graphs. Yet, the stress-based methods (PMDS, Maxent, SM) are the worst though they work quite well for a few mesh-like graphs (e.g., \emph{grid17}). Furthermore, traditional force-based methods (FR, FR-RVS, SFDP, Linlog, and FA2)  cannot correctly preserve the neighborhood for most graphs. In contrast, neighborhood embedding based methods (tsNET and  DRGraph) efficiently preserve neighborhoods for most graphs, whereas their performance with respect to the stress error is still far from t-FDP.
The reason is that these methods are designed for forming local clusters instead of capturing global structures.
Note that the NP1 and NP2 values of some graphs are quite small, no matter what the layout algorithm is. The graph \emph{lp\_ship04l} is an extreme case, where the NP1 value is 0.045. After carefully checking these graphs, we found that their intrinsic dimensions are very high, making a good projection nearly impossible, see Fig.~\ref{fig:Case}(c).
Since the parameters ``perplexity'' in tsNET and ``$k$-order nearest neighbors'' in DRGraph have a large impact on NP1 and NP2, we further investigated if the proper parameters can lead to better results. \yh{We found that t-FDP performs better than tsNET and DRGraph for most datasets and results in  higher mean values of SE, NP1 and NP2 among a set of parameters tested.} The detailed results can be found from the supplemental material.

The bottom row in Fig.~\ref{fig:heatmap} shows  the mean values of three metrics over 37 graphs that we were able to process with all layout methods. SM is definitely the best in SE and our t-FDP is the second best. Compared to FR and SFDP,
t-FDP performs similarly in SE but preserves more than 35\% and 15\% neighborhoods in NP1 and NP2, respectively, where the corresponding absolute mean differences are 0.22 and 0.10 (see the bottom row in Fig.~\ref{fig:heatmap})
Although Maxent performs only slightly worse than t-FDP for the stress error, it performs the worst for NP1 and NP2. %and  second-worst for edge crossings.
Similarly, PMDS is the second-worst for stress error, NP1, and NP2.
After carefully examining the layouts generated by Maxent and PMDS, we found that neither of them can efficiently handle graphs whose numbers of edges are larger than the numbers of nodes.
On the other hand, it is not surprising that SM is the best and LinLog is the worst for the stress error, since SM is designed for preserving distances while LinLog is for revealing clusters.

The boxplots in Fig.~\ref{fig:aQuality} summarize the EC and MA scores over 37 graphs that can be handled by all methods. Fig.~\ref{fig:aQuality}(a) shows that t-FDP performs similarly to the other methods w.r.t. EC, while PMDS and Maxent are slightly worse.
Fig.~\ref{fig:aQuality}(b) shows that Maxent performs the best in MA, followed by SM and t-FDP, while PMDS is the worst.
From these results, we conclude that the readability of  our t-FDP produced layouts is comparable and even superior to the existing methods.
%The other methods  perform similarly and are worse than t-FDP.}

Note that FR-RVS is significantly worse than FR w.r.t. SE and NP2, which contradicts the results from Gove~\cite{gove2019random}. After carefully checking our results, we found that the differences are due to initialization, \yh{where
    Gove~\cite{gove2019random} used  an initialization of  randomly distributed nodes  on a uniformly spaced disc. By using similar random initializations,} FR-RVS and FR perform similarly but the resulting layouts are worse than the ones generated by PMDS initialization. The full evaluation can be found in the supplemental material, where we also include the visual layouts generated by FR-RVS and t-FDP-RVS for comparison.

\vspace{1.5mm}
\noindent\textbf{Visual Results.}\
Fig.~\ref{fig:visual} shows the visual results of the six methods applied to four data sets: \emph{cluster}, \emph{bcspwr07}, \emph{3elt} and \emph{eva}.
We can see that t-FDP can characterize clusters as tsNET and DRGraph for the clustered graphs (see the top row), clearly revealing the global structures as FR and SM while  maintaining the neighborhood structures (see the second row), a good unfolding for mesh-like structures as FR (see the third row) and faithfully depicting multi-component structures (see the last row).
Although SM maintains a global structure by preserving pairwise distances, local structure preservation and clustering ability are the worst. FR and SFDP behave similarly to SM. tsNET and DRGraph seem to form clusters even when there are no cluster structures in the original graph, they also have difficulties in maintaining the global structure. \yh{For example, they squeeze
    the branches of the tree-like graph \emph{bcspwr07}  into a few small sub-clusters, which are overlapping (see the second row of Fig.~\ref{fig:visual}).}
Note that we did not employ any packing algorithm to arrange multiple components, whereas FR and SFDP tightly pack them in layout space.

%Specifically,  SM produces overlapping structures for the \emph{cluster} (top row) data set, FR and SFDP lead to partly mixed clusters, while they are clearly separated by tsNET, DRGraph and t-FDP layouts.
%%
%For \emph{bcspwr07} (middle), the outer vertices of SM, FR, and SFDP layouts tend to be placed close to each another, which lets local structures overlap and creates a poor neighborhood preservation (see also Fig.12).
%The vertices of tsNET and DRGraph layouts tend to be heavily contracted to various small clusters, making it hard to observe fine structures. Meanwhile, the long edges destroy the overall structure to some extent. t-FDP has a good balance between showing local neighbourhoods and backbone structures.
%%
%The layouts of \emph{3elt} all show three humps on the top-left for  FR, SFDP, SM, and t-FDP.  Only two humps can be seen for tsNET, DRGraph creates a more serious distortion of the overall structure.

\vspace{1.5mm}
\noindent\textbf{Runtime Performance.}
Zhu et al.~\cite{zhu2020drgraph} show in their experiments that only SFDP, PMDS, and DRGraph can handle graphs with millions of nodes. Besides these methods, we include Maxent for making
a comprehensive comparison with our t-FDP with regard to runtime performance.
%\od{while Fig.~\ref{fig:aQuality} reports the relative errors}.
For all methods, the runtime includes only the layout time without data processing steps such as data loading and layout initialization. Fig.~\ref{fig:Performance} reports the runtime for each graph.  %like the multi-level graph construction in SFDP.

In contrast to what was reported by Zhu et al.~\cite{zhu2020drgraph}, our implementation of PMDS is the fastest, because of its linear computational complexity. For small graphs with less than 1K nodes, all methods can be processed in less than a second, while the performances vary significantly for larger graphs.
SFDP, Maxent, and DRGraph have similar performances for graphs with more than 100K nodes, because of their similar computational complexity $O(n\log n)$ for computing repulsive forces.
The GPU version of the ibFFT based t-FDP is two orders of magnitude faster than these methods due to its ibFFT approximation.
For example, the GPU version of t-FDP model takes  30s for the \textit{com-LiveJournal} graph %(3,997,962 nodes and 34,681,189 edges)
(4 millions nodes and 35 million edges), while DRGraph, SFDP and Maxent require  2958s, 5745s and 11954s, respectively.

Based on the results of the five metrics and runtime performances,
we can conclude that our t-FDP model is able to generate high-quality layouts for most graphs while being extremely fast to compute.

% Except \textit{luxembourg_osm} is only about an order of magnitude over. But we can see from the layout quality results that our approach has significantly better neighbor preservation(NP1 = 0.898 and NP2 = 0.772) than others in the \textit{luxembourg_osm}, while the best of the others is SFDP(NP1 = 0.585 and NP2 = 0.532).
%\input{sections/extension}

%\vspace{-3mm}
\section{Extensions}
%respectively. One is to increase the weight of the repulsive t-force for enlarging the repulsion between nearby nodes, the other is to decrease the maximum magnitude of the repulsive t-force for moving connected nodes closer together in the visual representation. Initializing these two extensions with the results of the t-FDP model can quickly generate a variety of desired results.We demonstrate the effectiveness with three case studies.
Since our t-FDP model retains the flexibility and simplicity of traditional FDP, it inherits all possible extensions of these models, such as multi-level layout methods~\cite{hu2005efficient} or constrained layouts~\cite{dwyer2009scalable}.
Besides that, our t-FDP model can be further extended for better supporting  the interactive exploration of graphs by  globally and locally adjusting the repulsive t-force. \yh{Since our model is fast enough, these refinement extensions can be done with real-time interaction.}
%Since our model allows a very fast computation, a fluid exploration of parameter values  is ensured.
% Due to the natural flexibility of the FDP method, we can easily get a general layout and then re-layout it with different parameters to achieve different results.

\begin{figure}[!t]
	\centering
	\includegraphics[width=0.92\linewidth]{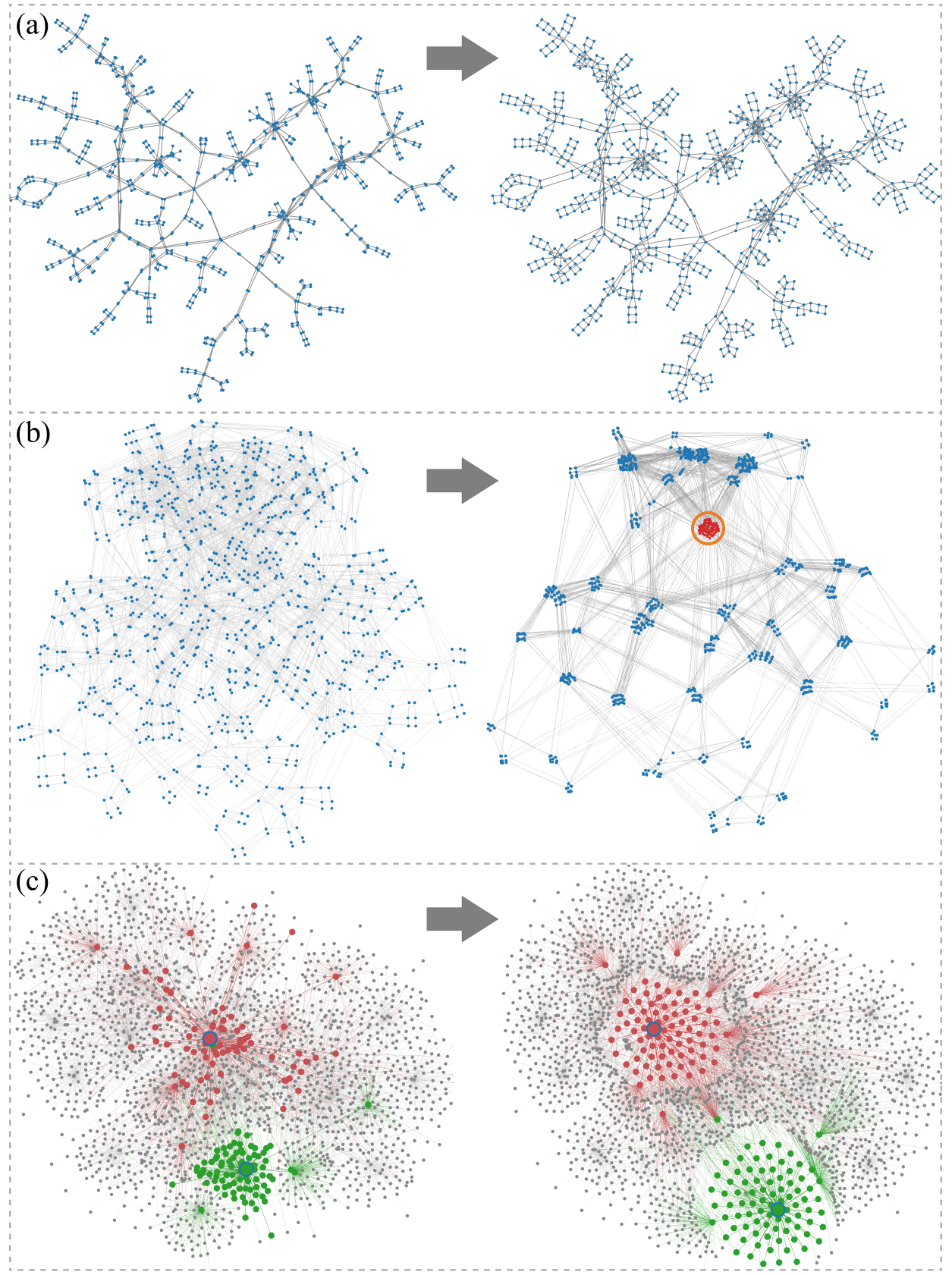}\vspace{-2mm}
	\caption{Refining t-FDP graph layouts by  applying a large repulsive force (a), and  a repulsive force with shorter range (b), and locally changing  attractive and repulsive forces (c). These  result in uniform distributed local neighborhoods (a), major clusters (b), and fisheye views (c).}
	\label{fig:Case}\vspace{-5mm}
\end{figure}

\vspace{1.5mm}
\noindent\textbf{Global Refinement.}
Users often want to explore different graph structures, such as detailed local neighborhoods or skeleton-like structures. Taking a t-FDP layout as  initialization, we can achieve such visualizations by re-applying t-FDP \yh{with repulsive t-forces of different values}. For example, using a large repulsive t-force will distribute nearby nodes evenly, resulting in detailed local neighborhoods. On the other hand, a small repulsive t-force will move connected nodes closer together and reveal major structures.

Fig.~\ref{fig:Case}(a) shows an example of applying a large repulsive force  to the layout of the graph \emph{qh882} (on the left), distributing the nodes at the tip of the branches evenly (right) and improving the NP1 score from 0.585 to 0.643. In both results,
the warping effect~\cite{yifanhuwarpingeffects} --that lets nodes in the periphery tend to be closer-- is greatly alleviated while unfolding all parts of the graph.
The example in Fig.~\ref{fig:Case}(b) is obtained in a reverse way: applying a repulsive force with a shorter range (larger $\gamma$) allows to display a clear skeleton structure of the graph \emph{cage8}.

%\od{To verify if the shown clusters are valid}, we computed the edge probabilities between nodes of the same clusters and nodes between different clusters and found that all revealed clusters are reasonable. For example,  the intra-cluster and inter-cluster edge probabilities of the cluster highlighted with the red point in the right of Fig.~\ref{fig:Case}(b) are 0.16 and 0.01, respectively.

\vspace{1.5mm}
\noindent\textbf{Local Refinement.}
During exploration, one major task is to find and examine the neighborhood of certain nodes. Although our t-FDP performs well in neighborhood preservation, it cannot ensure that all neighborhoods are always well preserved. The left in Fig.~\ref{fig:Case}(c) shows an example of  \emph{lp\_ship04l} graph, where some neighboring nodes (see the red and green ones) cannot be
%pulled together
placed properly
because of their own local clusters.
To alleviate this issue, we enhance the attractive forces between the focal nodes and their neighbors for pulling them together, while exerting repulsive forces to highlight them. Meanwhile, we exert large repulsive forces between other nodes for compressing the surrounding area.
In doing so, a fisheye-like visualization is generated.
The right in Fig.~\ref{fig:Case}(c) shows the result, where most neighborhoods of the red and green nodes are clearly revealed. Note that a few nodes are still not pulled together because of their own local clusters.

% \begin{figure}[htp]
%     \centering
%     \includegraphics[width=0.999\linewidth]{Case.png}
%     \caption{}
%     \label{fig:Case}
% \end{figure}

% Perhaps it would be better to write it in categories by method. Like following:

% SM and FR can obtain good distance preservation and global structure on most of the graphs. SM performs best on grid17 with uniform node distribution and tidy grids, because the graph has simple local structure. FR causes peripheral effect\cite{hu2005efficient}, which vertices in the peripheral tends to be closer to each other than those in the center. For \textit{qh882}, \textit{btree} and \textit{USPowerGrid}, SM and FR lead to many parts of the graph overlap (crowded problem like MDS in DR domain), and the numerical results show the poor neighbor preservation of SM and FR on these graphs. Synthesized graphs with clusters(\textit{cluster800}) and hierarchical clusters(\textit{clusterHiera1600}) shows poor ability of SM and FR to show clusters.

\section{Conclusion and Future Work}
We present t-FDP, a novel FDP model for graph placement. It is based on the observation that  existing FDP models cannot properly capture local neighborhoods, especially due to the large contact forces when two nodes overlap.

Therefore, we devise a new short-range force based on the t-distribution: t-force. It  has a defined upper bound, behaves similar to existing power functions based attractive forces at long-range and repulsive forces at short-range. Furthermore, we adapt the FFT based approximation strategy used for t-SNE to accelerate the computation of the repulsive force.
We quantitatively compare our t-FDP model with different state-of-the-art layout methods, showing that the t-FDP based on the FFT approximation outperforms them in most cases and is one magnitude faster on CPU and two orders faster using a GPU.
Lastly, we demonstrate the usefulness of t-FDP in exploring different graph structures.

Our approach still has certain limitations, which we would like to address in the future:
First, t-FDP performs slightly worse than stress models in distance preservation, although it shows better results than tsNET and DRGraph. We therefore plan to explore the possibility of improving its long-range forces. %Second, t-FDP also cannot clearly reveal major structures in large social networks (as most existing methods) because of too many edges. One possible way is to first extract the backbone structures (see Fig.~\ref{fig:Case}(b)) and then use them to drive the layout.
Second, t-FDP replaces a single parameter for balancing attractive and repulsive forces acting on each node by three parameters $\alpha$, $\beta$ and $\gamma$,  which might result in additional complications for users. On the other hand, these parameters provide freedom for adapting the method to different tasks~\cite{lee2006task}, therefore, we like to explore automated parameter tuning methods.
Finally, we plan to investigate other possible forms of  short-range forces to further improve layout quality and explore their applications in dimensionality reduction. %~\cite{van2009learning}.

\section*{Acknowledgments}
The authors like to thank the anonymous reviewers for their valuable input, this work was supported by the grants of the National Key Research \& Development Plan of China (2019YFB1704201) and NSFC (62132017, 62141217), as well as
by the Deutsche Forschungsgemeinschaft (DFG, German Research Foundation) under Germany's Excellence Strategy - EXC 2117 - 422037984.

\bibliographystyle{IEEEtran}

\bibliography{tFDP}

\begin{IEEEbiography}[{\includegraphics[width=1in,height=1.25in,clip,keepaspectratio]{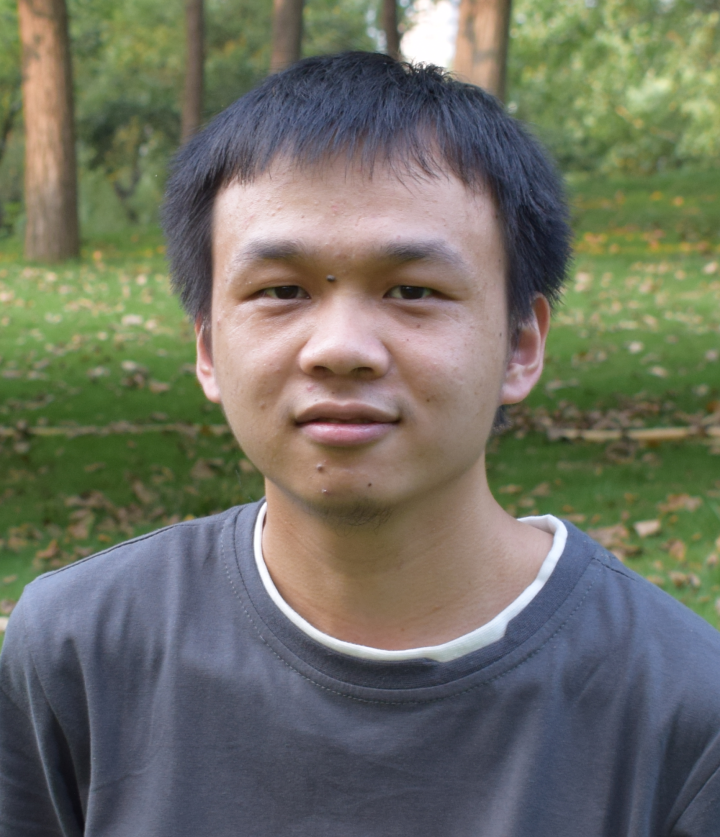}}]{Fahai Zhong}
	is a third-year Master student in the School of Computer Science and Technology, at Shandong University. His research interests include graph visualization and dimensional reduction.
\end{IEEEbiography}
% \vspace {-9mm}

\begin{IEEEbiography}[{\includegraphics[width=1in,height=1.25in,clip,keepaspectratio]{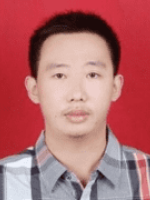}}]{Mingliang Xue}
	is a fifth-year Ph.D. student in the School of Computer Science and Technology, Shandong University. His research interests include graph visualization and dimensional reduction.
\end{IEEEbiography}
% \vspace {-9mm}
\begin{IEEEbiography}[{\includegraphics[width=1in,height=1.25in,clip,keepaspectratio]{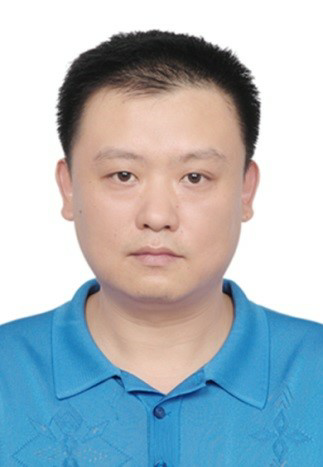}}]{Jian Zhang}
	received his PhD degree in Applied Mathematics from the University of Minnesota in 2005.
	After a postdoc at Pennsylvania State University, he is now a professor in the Computer Network Information Center, Chinese Academy of Sciences (CAS).
	His current research interests include scientific computing and scientific visualization.
\end{IEEEbiography}

\begin{IEEEbiography}[{\includegraphics[width=1in,height=1.25in,clip,keepaspectratio]{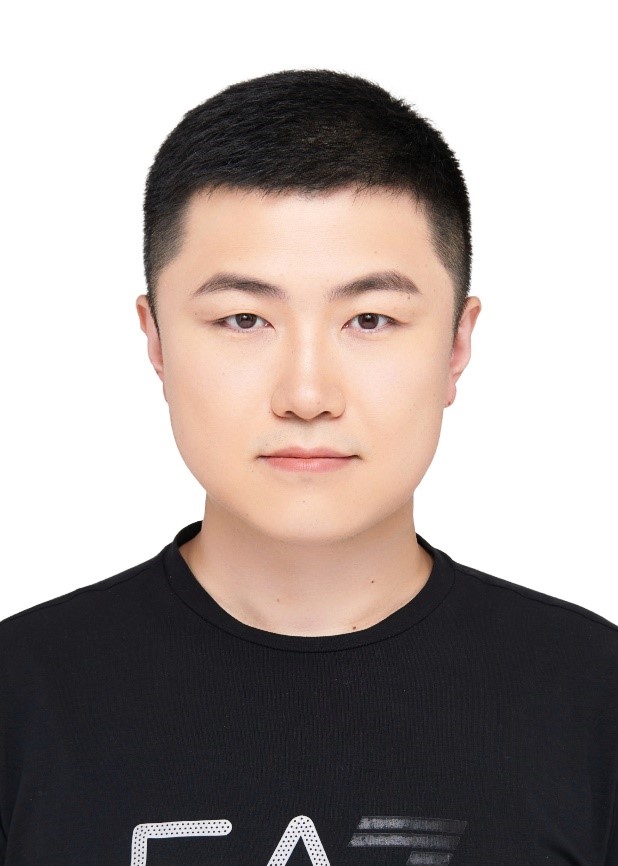}}]{Fan Zhang}
	received the B.Eng. and Ph.D. degrees from the School of Computer Science and Technology, Shandong University, China, in 2009 and 2015, respectively. He is currently an associate professor at the School of Computer Science and Technology, Shandong Technology and Business University, China. His research interests include computer graphics, computer vision, and artificial intelligence.
\end{IEEEbiography}

% \vspace {-9mm}
\begin{IEEEbiography}[{\includegraphics[width=1in,height=1.25in,clip,keepaspectratio]{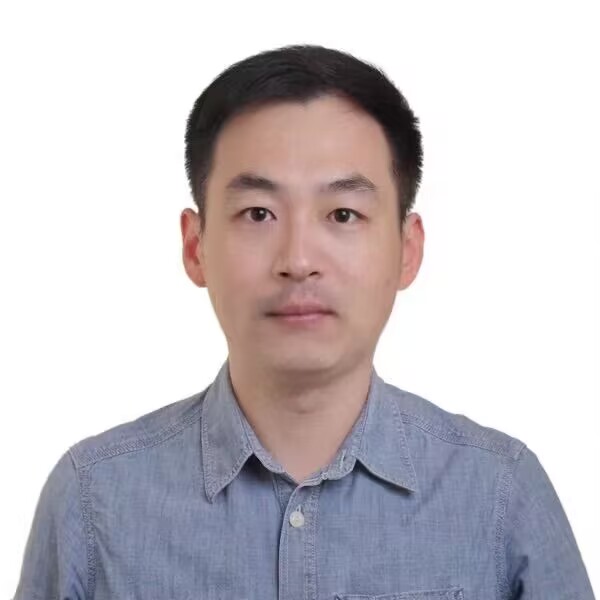}}]{Rui Ban}
	is a senior engineer at Intelligent Network Design Institute of China Information Technology Designing Consulting Institute Co., Ltd. His research interests include big data visualization and visual analytics.
\end{IEEEbiography}
% \vspace {-9mm}
\begin{IEEEbiography}[{\includegraphics[width=1in,height=1.25in,clip,keepaspectratio]{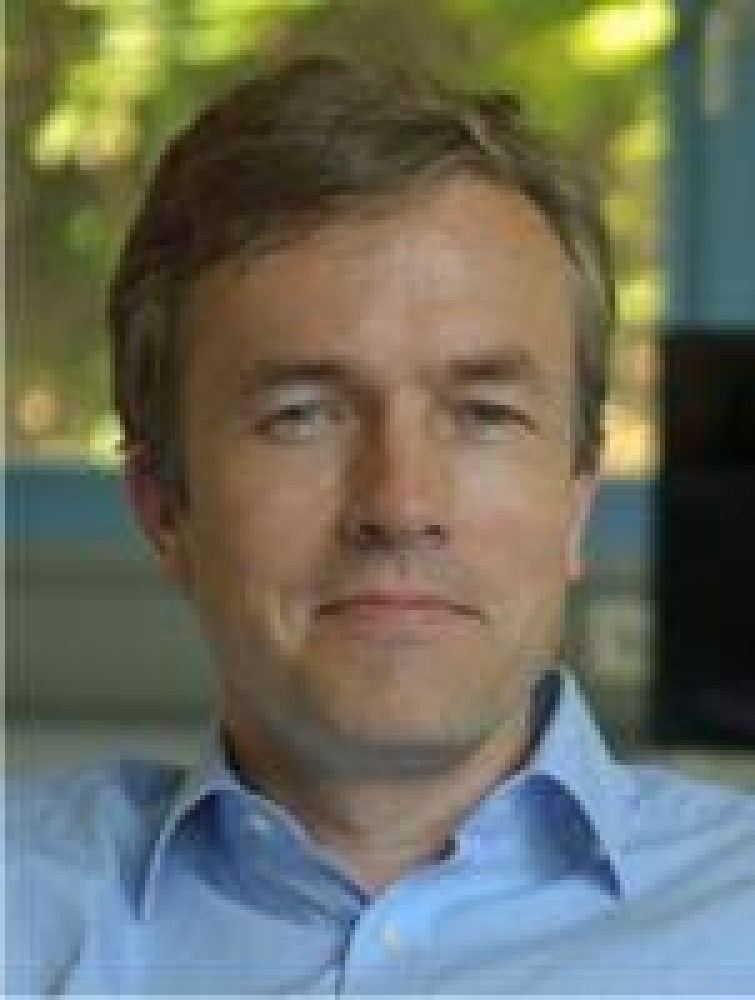}}]{Oliver Deussen}
	graduated at Karlsruhe Institute of Technology and is professor at University of Konstanz (Germany) and visiting professor at the Chinese Academy of Science in Shenzhen.
	He served as Co-Editor in Chief of Computer Graphics Forum and was President of the  Eurographics Association.
	His areas of interest are modeling and rendering of complex biological systems, non-photorealistic rendering as well as Information Visualization.
\end{IEEEbiography}
% \vspace {-9mm}
\begin{IEEEbiography}[{\includegraphics[width=1in,height=1.25in,clip,keepaspectratio]{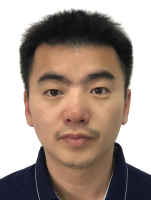}}]{Yunhai Wang}
	is professor in School of Computer Science and Technology at Shandong University.
	He serves as the associate editor of Computer Graphics Forum.
	His interests include scientific visualization, information visualization and computer graphics.
\end{IEEEbiography}
\end{document}